\documentclass[]{svmult}

\usepackage{mathptmx}       
\usepackage{helvet}         
\usepackage{courier}        
\usepackage{amsfonts}
\usepackage{epsf,epsfig}
\usepackage{amsmath}
\usepackage{fancybox}
\usepackage{amssymb}

\usepackage{type1cm}        
\usepackage{makeidx}         
\usepackage{graphicx}        
\usepackage{multicol}        
\usepackage[bottom]{footmisc}
\input{Qcircuit}

\newcommand{\norm}[1]{\left\lvert#1\right\rvert}
\newcommand{\vnorm}[1]{\big\lVert#1\big\rVert}

\newcommand{\braket}[2]{\langle #1 | #2\rangle}    
\newcommand{\ketbra}[2]{\ket{#1}\!\bra{#2}}        
\newcommand{\pr}[1]{\ketbra{#1}{#1}}        
\newcommand{\density}[1]{\ketbra{#1}{#1}}     
\newcommand{\Ket}[1]{\left|#1\right\rangle}        

\renewcommand{\mod}{\hspace{1mm}\textnormal{mod}\hspace{1mm}}

\renewcommand{\Pr}{\text{Pr}}

\renewcommand{\P}{\text{{\it P}}}

\newcommand{\Z}{\mathbb{Z}}

\newcommand{\x}{\mathbf{x}}
\newcommand{\y}{\mathbf{y}}

\newcommand{\s}{\mathbf{s}}

\newcommand{\0}{\mathbf{0}}
\newcommand{\bee}{\mathbf{b}}
\newcommand{\ay}{\mathbf{a}}

\newcommand{\size}[1]{\left|#1\right|}             

\newcommand{\half}{\frac{1}{2}}

\newcommand{\oort}{\frac{1}{\sqrt{2}}}
\newcommand{\tbt}[4]{\begin{pmatrix}#1&#2\\#3&#4\end{pmatrix}}

\newcommand{\be}{\begin{equation}}
\newcommand{\ee}{\end{equation}}
\newcommand{\bea}{\begin{eqnarray}}
\newcommand{\eea}{\end{eqnarray}}
\newcommand{\ths}{\thickspace}
\newcommand{\pri}{^\prime}
\newcommand{\ddt}{\frac{\text{d}}{\text{dt}}}
\newcommand{\AND}{\text{{\sc and}}}
\newcommand{\OR}{\text{{\sc or}}}
\newcommand{\NAND}{\text{{\sc nand}}}
\newcommand{\NOT}{\text{{\sc not}}}

\makeindex             

\title*{Algorithms for Quantum Computers}
\author{Jamie Smith and Michele Mosca}
\institute{Jamie Smith \at Institute for Quantum Computing and Dept. of Combinatorics \& Optimization \\ University of Waterloo, \\with support from the Natural Sciences and Engineering Research Council of Canada\\\email{ja5smith@iqc.ca}
\and Michele Mosca \at Institute for Quantum Computing and Dept. of Combinatorics \& Optimization \\ University of Waterloo and St. Jerome's University, \\ and Perimeter Institute for Theoretical Physics,\\ with support from the Government of Canada,  Ontario-MRI, NSERC, QuantumWorks, MITACS, CIFAR, CRC, ORF, and DTO-ARO\\\email{mmosca@iqc.ca}}

\begin{document}
\maketitle
\section{Introduction}

Quantum computing is a new computational paradigm created by reformulating information and computation in a quantum mechanical framework \cite{Fey82, Deu85}. Since the laws of physics appear to be quantum mechanical, this is the most relevant framework to consider when considering the fundamental limitations of information processing. Furthermore, in recent decades we have seen a major shift from just observing quantum phenomena to actually controlling quantum mechanical systems. We have seen the communication of quantum information over long distances, the ``teleportation'' of quantum information, and the encoding and manipulation of quantum information in many different physical media. We still appear to be a long way from the implementation of a large-scale quantum computer, however it is a serious goal of many of the world's leading physicists, and progress continues at a fast pace.

In parallel with the broad and aggressive program to control quantum mechanical systems with increased precision, and to control and interact a larger number of subsystems, researchers have also been aggressively pushing the boundaries of what useful tasks one could perform with quantum mechanical devices. These include improved metrology, quantum communication and cryptography, and the implementation of large-scale quantum algorithms.

It was known very early on \cite{Deu85} that quantum algorithms cannot compute functions that are not computable by classical computers, however they might be able to efficiently compute functions that are not efficiently computable on a classical computer. Or, at the very least, quantum algorithms might be able to provide some sort of speed-up over the best possible or best known classical algorithms for a specific problem.

The purpose of this paper is to survey the field of quantum algorithms, which has grown tremendously since Shor's breakthrough algorithms \cite{Sho94, Sho97} over 15 years ago.
Much of the work in quantum algorithms is now textbook material (e.g. \cite{NC00, Hir01, KSV02, KLM07, Mer07} ), and we will only briefly mention these examples in order to provide a broad overview.  Other parts of this survey, in particular, sections \ref{sec:quantum-walks} and \ref{sec:tensor-networks}, give a more detailed description of some more recent work.

We organized this survey according to an underlying tool or approach taken, and include some basic applications and specific examples, and relevant comparisons with classical algorithms.

In section \ref{sec:quantum-fourier-transform} we begin with algorithms one can naturally consider to be based on a quantum Fourier transform, which includes the famous factoring and discrete logarithm algorithms of Peter Shor \cite{Sho94, Sho97}. Since this topic is covered in several textbooks and recent surveys, we will only briefly survey this topic. We could have added several other sections on algorithms for generalizations of these problems, including several cases of the non-Abelian hidden subgroup problem, and hidden lattice problems over the reals (which have important applications in number theory), however these are covered in the recent survey \cite{Mos09} and in substantial detail in \cite{CD09}.

We continue in section \ref{sec:amplitude-amplification} with a brief review of classic results on quantum searching and counting, and more generally amplitude amplification and amplitude estimation.

In section \ref{sec:quantum-walks} we discuss algorithms based on quantum walks. We will not cover the related topic of adiabatic algorithms, which was briefly summarized in \cite{Mos09}; a broader survey of this and related techniques (``quantum annealing'') can be found in \cite{DC08}.

We conclude with section \ref{sec:tensor-networks} on algorithms based on the evaluation of the trace of an operator, also referred to as the evaluation of a tensor network, and which has applications such as the approximation of the Tutte polynomial.

The field of quantum algorithms has grown tremendously since the seminal work in the mid-1990s, and a full detailed survey would simply be infeasible for one article. We could have added several other sections. One major omission is the development of algorithms for simulating quantum mechanical systems, which was Feynman's original motivation for proposing a quantum computer. This field was briefly surveyed in \cite{Mos09}, with emphasis on the recent results in \cite{BACS07} (more recent developments can be found in \cite{WBHS08}). This remains an area worthy of a comprehensive survey; like the many other areas we have tried to survey, it is difficult because it is still an active area of research. It is also an especially important area because these algorithms tend to offer a fully exponential speed-up over classical algorithms, and thus are likely to be among the first quantum algorithms to be implemented that will offer a speed-up over the fastest available classical computers.

Lastly, one could also write a survey of quantum algorithms for intrinsically quantum information problems, like entanglement concentration, or quantum data compression.  We do not cover this topic in this article, though there is a very brief survey in \cite{Mos09}.

One can find a rather comprehensive list of the known quantum algorithms (up to mid 2008) in Stephen Jordan's PhD thesis \cite{Jor08}.
We hope this survey complements some of the other recent surveys in providing a reasonably detailed overview of the current state of the art in quantum algorithms.

\section{Algorithms based on the Quantum Fourier transform}\label{sec:quantum-fourier-transform}

The early line of quantum algorithms was developed in the ``black-box'' or ``oracle'' framework.
In this framework part of the input is a black-box that implements a function $f(x)$, and the only way to extract information about $f$ is to evaluate it on inputs $x$. These early algorithms
 used a special case of quantum Fourier transform, the Hadamard gate, in order solve the given problem with fewer black-box evaluations of $f$ than a classical algorithm would require.

Deutsch \cite{Deu85} formulated the problem of deciding whether a function $f: \{0,1\} \rightarrow \{0,1\}$ was constant or not. Suppose one has access to a black-box that implements $f$ reversibly by mapping $x,0 \mapsto x,f(x)$; let us further assume that the black box in fact implements a unitary transformation $U_f$ that maps $\ket{x} \ket{0} \mapsto \ket{x} \ket{f(x)}$. Deutsch's problem is to output ``constant'' if $f(0)=f(1)$ and to output ``balanced'' if $f(0) \neq f(1)$, given a black-box for evaluating $f$. In other words determine $f(0)\oplus f(1)$ (where $\oplus$ denotes addition modulo $2$). Outcome ``0'' means $f$ is constant and ``1'' means $f$ is not constant.

A classical algorithm would need to evaluate $f$ twice in order to solve this problem. A quantum algorithm can apply $U_f$ only once to create \[ \frac{1}{\sqrt{2}} \ket{0} \ket{f(0)} + \frac{1}{\sqrt{2}} \ket{1} \ket{f(1)} .\]

Note that if $f(0) = f(1)$, then applying the Hadamard gate to the first register yields $\ket{0}$ with probability $1$, and if $f(0) \neq f(1)$, then applying the Hadamard gate to the first register and ignoring the second register leaves the first register in the state $\ket{1}$ with probability $\frac{1}{2}$; thus a result of $\ket{1}$ could only occur if $f(0) \neq f(1)$.

As an aside, let us note that in general, given
\[ \frac{1}{\sqrt{2}} \ket{0} \ket{\psi_0} + \frac{1}{\sqrt{2}} \ket{1} \ket{\psi_1} \] applying the Hadamard gate to the first qubit and measuring it will yield ``0'' with probability $\frac{1}{2} + Re( \bra{\psi_0} \psi_1 \rangle)$ ; this ``Hadamard test'' is discussed in more detail in section \ref{sec:tensor-networks}.

In Deutsch's case, measuring a ``1'' meant $f(0) \neq f(1)$ with certainty, and a ``0'' was an inconclusive result. Even though it wasn't perfect, it still was something that couldn't be done with a classical algorithm. The algorithm can be made exact \cite{CEMM98} (i.e. one that outputs the correct answer with probability $1$) if one assumes further that $U_f$ maps $\ket{x} \ket{b} \mapsto \ket{x} \ket{b \oplus f(x)}$, for $b \in \{0,1\}$, and one sets the second qubit to $ \frac{1}{\sqrt{2}} \ket{0} - \frac{1}{\sqrt{2}} \ket{1}$. Then $U_f$ maps
\[ \left( \frac{\ket{0} + \ket{1}}{\sqrt{2}} \right) \left( \frac{\ket{0} - \ket{1}}{\sqrt{2}} \right) \mapsto
(-1)^{f(0)}\left( \frac{\ket{0} + (-1)^{f(0) \oplus f(1)} \ket{1}}{\sqrt{2}} \right) \left( \frac{\ket{0} - \ket{1}}{\sqrt{2}} \right) .\]
Thus a Hadamard gate on the first qubit yields the result
\[ (-1)^{f(0)} \ket{f(0) \oplus f(1)} \left( \frac{\ket{0} - \ket{1}}{\sqrt{2}} \right) \] and measuring the first register yields the correct answer with certainty.

The general idea behind the early quantum algorithms was to compute a black-box function $f$ on a superposition of inputs, and then extract a global property of $f$ by applying a quantum transformation to the input register before measuring it.  We usually assume we have access to a black-box that implements
\[ U_f : \ket{\x} \ket{\bee} \mapsto \ket{\x} \ket{\bee \oplus f(\x)} \] or in some other form where the input value $\x$ is kept intact and the second register is shifted by $f(\x)$ is some reversible way.

Deutsch and Jozsa  \cite{DJ92} used this approach to get an exact algorithm that decides whether $f:\{0,1\}^n \mapsto \{0,1\}$ is constant or ``balanced'' (i.e. $|f^{-1}(0)| = |f^{-1}(1)|$), with a promise that one of these two cases holds.  Their algorithm evaluated $f$ only twice, while classically any exact algorithm would require $2^{n-1}+1$ queries in the worst-case.
Bernstein and Vazirani \cite{BV97} defined a specific class of such functions $f_{\ay }:\x \mapsto \ay \cdot \x$, for any $\ay \in \{0,1\}^n$, and showed how the same algorithm that solves the Deutsch-Jozsa problem allows one to determine $\ay$ with two evaluations of $f_{\ay }$ while a classical algorithm requires $n$ evaluations. (Both of these algorithms can be done with one query if we have $\ket{\x} \ket{b} \mapsto \ket{\x}\ket{b \oplus f(\x)}$.) They further showed how a related ``recursive Fourier sampling'' problem could be solved super-polynomially faster on a quantum computer. Simon \cite{Sim94} later built on these tools to develop a black-box quantum algorithm that was exponentially faster than any classical algorithm for finding a hidden string $\s \in \{0,1\}^n$ that is encoded in a function $f :\{0,1\}^n \rightarrow \{0,1\}^n$ with the property that $f(\x) = f(\y)$ if and only if $\x = \y \oplus \s$.

Shor \cite{Sho94, Sho97}  built on these black-box results to find an efficient algorithm for finding the order of an element in the multiplicative group of integers modulo $N$ (which implies an efficient classical algorithm for factoring $N$) and for solving the discrete logarithm problem in the multiplicative group of integers modulo a large prime $p$. Since the most widely used public key cryptography schemes at the time relied on the difficulty of integer factorization, and others relied on the difficulty of the discrete logarithm problem, these results had very serious practical implications. Shor's algorithms straightforwardly apply to black-box groups, and thus permit finding orders and discrete logarithms in any group that is reasonably presented, including the additive group of points on elliptic curves, which is currently one of the most widely used public key cryptography schemes (see e.g. \cite{MVV96}).

Researchers tried to understand the full implications and applications of Shor's technique, and a number of generalizations were soon formulated (e.g. \cite{BL95, Gri97}). One can phrase Simon's algorithm, Shor's algorithm, and the various generalizations that soon followed as special cases of the {\it hidden subgroup problem}.
Consider a finitely generated Abelian group $G$, and a hidden subgroup $K$ that is defined by a function $f : G \rightarrow X$ (for some finite set $X$) with the property that $f(x) = f(y)$ if and only if $x-y \in K$ (we use additive notation, without loss of generality). In other words, $f$ is constant on cosets of $K$ and distinct on different cosets of $G$.
In the case of Simon's algorithm, $G = \Z_2^n$ and $K = \{\0, \s\}$. In the case of Shor's order-finding algorithm, $G = \Z$ and $K = r\Z$ where $r$ is the unknown order of the element. Other examples and how they fit in the hidden subgroup paradigm are given in \cite{Mos08}.

Soon after, Kitaev \cite{Kit95} solved a problem he called the {\it Abelian stabilizer problem} using an approach that seemed different from Shor's algorithm, one based in eigenvalue estimation. Eigenvalue estimation is in fact an algorithm of independent interest for the purpose of studying quantum mechanical systems. The Abelian stabilizer problem is also a special case of the hidden subgroup problem. Kitaev's idea was to turn the problem into one of estimating eigenvalues of unitary operators. In the language of the hidden subgroup problem, the unitary operators were shift operators of the form $f(x) \mapsto f(x + y)$. By encoding the eigenvalues as relative phase shifts, he turned the problem into a phase estimation problem.

The Simon/Shor approach for solving the hidden subgroup problem is to first compute $\sum_x \ket{x} \ket{f(x)}$.
In the case of finite groups $G$, one can sum over all the elements of $G$, otherwise one can sum over a sufficiently large subset of $G$. For example, if $G=\Z$, and $f(x) = a^x \mod N$ for some large integer $N$, we first compute $\sum_{x=0}^{2^n-1} \ket{x} \ket{a^x}$, where $2^n > N^2$ (we omit the ``$\mod N$'' for simplicity).
If $r$ is the order of $a$ (i.e. $r$ is the smallest positive integer such that $a^r \equiv 1$) then every value $x$ of the form $x = y+kr$ gets mapped to $a^y$. Thus we can rewrite the above state as
\begin{eqnarray} \label{shor-eqn}
 \sum_{x=0}^{2^n-1} \ket{x} \ket{a^x} = \sum_{y=0}^{r-1} (\sum_{j} \ket{y + jr}) \ket{a^y}
  \end{eqnarray}
  where each value of $a^y$ in this range is distinct.
Tracing out the second register we thus are left with a state of the form \[ \sum_{j} \ket{y + jr} \] for a random $y$ and where $j$ goes from $0$ to $\lfloor (2^n-1)/r \rfloor$. We loosely refer to this state as a ``periodic'' state with period $r$. We can use the inverse of the quantum Fourier transform (or the quantum Fourier transform) to map this state to a state of the form $\sum_x \alpha_x \ket{x}$, where the amplitudes are biased towards values of $x$ such that $x/2^n \approx k/r$. With probability at least $\frac{4}{\pi^2}$ we obtain an $x$ such that $|x/2^n - k/r| \leq 1/2^{n+1}$. One can then use the continued fractions algorithm to find $k/r$ (in lowest terms) and thus find $r$ with high probability. It is important to note that the continued fractions algorithm is not needed for many of the other cases of the Abelian hidden subgroup considered, such as Simon's algorithm or the discrete logarithm algorithm when the order of the group is already known.

In contrast, Kitaev's approach for this special case was to consider the map $U_a : \ket{b} \mapsto \ket{b a^{x}}$. It has eigenvalues of the form $e^{2 \pi i k/r}$, and the state $\ket{1}$ satisfies
\[ \ket{1} = \frac{1}{\sqrt{r}} \sum_{k=0}^{r-1} \ket{\psi_k} \] where $\ket{\psi_k}$ is an eigenvector with eigenvalue $e^{2 \pi i k/r}$:
 \[ U_a : \ket{\psi_k} \mapsto e^{2 \pi i kx/r} \ket{\psi_k} .\]
If we consider the controlled-$U_a$, denoted $c-U_a$, which maps $\ket{0} \ket{b} \mapsto \ket{0} \ket{b}$ and  $\ket{1} \ket{b} \mapsto \ket{1} \ket{ba}$, and if we apply it to the state $(\ket{0} + \ket{1}) \ket{\psi_k}$ we get
\[ (\ket{0} + \ket{1}) \ket{\psi_k} \mapsto (\ket{0} + e^{2 \pi i k/r} \ket{1}) \ket{\psi_k}  .\]
In other words, the eigenvalue becomes a relative phase, and thus we can reduce eigenvalue estimation to phase estimation.
Furthermore, since we can efficiently compute $a^{2^j}$ by performing $j$ multiplications modulo $N$, one can also efficiently implement $c-U_{a^{2^j}}$ and thus easily obtain the qubit $\ket{0} + e^{2 \pi i 2^j (k/r)} \ket{1}$ for integer values of $j$ without performing $c-U_a$ a total of $2^j$ times.
Kitaev developed an efficient ad hoc phase estimation scheme in order to estimate $k/r$ to high precision, and this phase estimation scheme could be optimized further \cite{CEMM98}. In particular, one can create the state
\[ (\ket{0}+\ket{1})^{n} \ket{1} = \sum_{x=0}^{2^n-1} \ket{x} \ket{1} =  \sum_{k=0}^{r-1} \sum_{x=0}^{2^n-1} \ket{x} \ket{\psi_k} \]
 (we use the standard binary encoding of the integers $x \in \{0,1, \ldots, 2^n-1\}$ as bits strings of length $n$) the apply the $c-U_{a^{2^j}}$ using the $(n-j)$th bit as the control bit, and using the second register (initialized in $\ket{1} = \sum_k \ket{\psi_k}$) as the target register, for $j=0,1,2, \ldots, n-1$, to create
 \begin{eqnarray} \label{kit-eqn}
   & \sum_k (\ket{0}+ e^{2 \pi i 2^{n-1}\frac{k}{r}} \ket{1}) \cdots (\ket{0}+ e^{2 \pi i 2\frac{k}{r}} \ket{1})(\ket{0}+ e^{2 \pi i \frac{k}{r}} \ket{1}) \ket{\psi_k} \\
  & =  \sum_{x=0}^{2^n-1} \ket{x} \ket{a^x} =  \sum_{k=0}^{r-1} \sum_{x=0}^{2^n-1} e^{2 \pi i x \frac{k}{r}} \ket{x} \ket{\psi_k} .\end{eqnarray}

If we ignore or discard the second register, we are left with a state of the form $\sum_{x=0}^{2^n-1} e^{2 \pi i x k/r} \ket{x}$ for a random value of $k \in \{0,1, \ldots, r-1\}$.
 The inverse quantum Fourier transformation maps this state to a state $\sum_y \alpha_y \ket{y}$ where most of weight of the amplitudes is near values of $y$ such that $y/2^n \approx k/r$ for some integer $k$. More specifically
 $|y/2^n - k/r| \leq 1/2^{n+1}$ with probability at least $4/\pi^2$; furthermore
 $|y/2^n - k/r| \leq 1/2^{n}$ with probability at least $8/\pi^2$. As in the case of Shor's algorithm, one can use the continued fractions algorithm to determine $k/r$ (in lowest terms) and thus determine $r$ with high probability.

It was noted \cite{CEMM98} that this modified eigenvalue estimation algorithm for order-finding was essentially equivalent to Shor's period-finding algorithm for order-finding. This can be seen by noting that we have the same state in Equation \ref{shor-eqn} and Equation \ref{kit-eqn}, and in both cases we discard the second register and apply an inverse Fourier transform to the first register. The only difference is the basis in which the second register is mathematically analyzed.

The most obvious direction in which to try to generalized the Abelian hidden subgroup algorithm is to solve instances of the hidden subgroup problem for non-Abelian groups. This includes, for example, the graph automorphism problem (which corresponds to finding a hidden subgroup of the symmetric group). There has been non-trivial, but limited, progress in this direction, using a variety of algorithmic tools, such as sieving, ``pretty good measurements'' and other group theoretic approaches. Other generalizations include the hidden shift problem and its generalizations, hidden lattice problems on real lattices (which has important applications in computational number theory and computationally secure cryptography), and hidden non-linear structures.
These topics and techniques would take several dozen pages just to summarize, so we refer the reader to \cite{Mos09} or \cite{CD09}, and leave more room to summarize other important topics.

\section{Amplitude Amplification and Estimation}\label{sec:amplitude-amplification}

A very general problem for which quantum algorithms offer an improvement is that of searching for a solution to a computational problem in the case that a solution $x$ can be easily verified. We can phrase such a general problem as finding a solution $x$ to the equation $f(x)=1$, given a means for evaluating the function $f$. We can further assume that $f:\{0,1\}^n \mapsto \{0,1\}$.

The problems from the previous section can be rephrased in this form (or a small number of instances of problems of this form), and the quantum algorithms for these problems exploit some non-trivial algebraic structure in the function $f$ in order to solve the problems superpolynomially faster than the best possible or best known classical algorithms.  Quantum computers also allow a more modest speed-up (up to quadratic) for searching for a solution to a function $f$ without any particular structure. This includes, for example, searching for solutions to $NP$-complete problems.

Note that classically, given a means for guessing a solution $\x$ with probability $p$, one could amplify this success probability by repeating many times, and after a number of guess in $O(1/p)$, the probability of finding a solution is in $\Omega(1)$. Note that the quantum implementation of an algorithm that produces a solution with probability $p$ will produce a solution with probability amplitude $\sqrt{p}$. The idea behind quantum searching is to somehow amplify this probability to be close to $1$ using only $O(1/\sqrt{p})$ guesses and other steps.

Lov Grover \cite{Gro96} found precisely such an algorithm, and this algorithm was analyzed in detail and generalized \cite{BBHT98, BH97, Gro98, BHMT00} to what is known as ``amplitude amplification.''
Any procedure for guessing a solution with probability $p$ can be (with modest overhead) turned into a unitary operator $A$ that maps $\ket{\0} \mapsto \sqrt{p} \ket{\psi_1} + \sqrt{1-p} \ket{\psi_0}$, where $\0 = 00 \ldots 0$, $\ket{\psi_1}$ is a superposition of states encoding solutions $\x$ to $f(\x)=1$ (the states could in general encode $\x$ followed by other ``junk'' information) and $\ket{\psi_0}$ is a superposition of states encoding values of $\x$ that are not solutions.

One can then define the quantum search iterate (or ``Grover iterate'') to be
\[ Q = -A U_0 A^{-1} U_f \] where
$U_f : \ket{\x} \mapsto (-1)^{f(\x)} \ket{\x}$, and $U_0 = I - 2\ket{\0} \bra{\0}$ (in other words, maps $\ket{\0} \mapsto -\ket{\0}$ and $\ket{\x} \mapsto \ket{\x}$ for any $\x \neq \0$). Here we are for simplicity assuming there are no ``junk'' bits in the unitary computation of $f$ by $A$.
Any such junk information can either be ``uncomputed'' and reset to all $0$s, or even ignored (letting $U_f$ act only on the bits encoding $\x$ and applying the identity to the junk bits).

This algorithm is analyzed thoroughly in the literature and in textbooks, so we only summarize the main results and ideas that might help understand the later sections on searching via quantum walk.

If one applies $Q$ a total of $k$ times to the input state $A \ket{00 \ldots 0} = \sin(\theta) \ket{\psi_1} + \cos(\theta) \ket{\psi_0}$, where $\sqrt{p} = \sin(\theta)$, then one obtains
\[ Q^k A \ket{00 \ldots 0} = \sin((2k+1) \theta) \ket{\psi_1} + \cos((2k+1)\theta) \ket{\psi_0} .\]

This implies that with $k \approx \frac{\pi}{4 \sqrt{p}}$ one obtains $\ket{\psi_1}$ with probability amplitude close to $1$, and thus measuring the register will yield a solution to $f(x)=1$ with high probability. This is quadratically better than what could be achieved by preparing and measuring $A \ket{\0}$ until a solution is found.

One application of such a generic amplitude amplification method is for searching.
One can also apply this technique to approximately count \cite{BHT97} the number of solutions to $f(x)=1$, and more generally to estimate the amplitude \cite{BHMT00} with which a general operator $A$ produces a solution to $f(x)=1$ (in other words, the transition amplitude from one recognizable subspace to another).

There are a variety of other applications of amplitude amplification that cleverly incorporate amplitude amplification in a non-obvious way into a larger algorithm. Some of these examples are discussed in \cite{Mos09} and most are summarized at \cite{Jor08}.

Since there are some connections to some of the more recent tools developed in quantum algorithms, we will briefly explain how amplitude estimation works.

Consider any unitary operator $A$ that maps some known input state, say $\ket{\0} = \ket{00 \ldots 0}$, to a superposition $\sin(\theta) \ket{\psi_1} + \cos(\theta) \ket{\psi_0}$, $0 \leq \theta \leq \pi/2$, where $\ket{\psi_1}$ is a normalized superposition of ``good'' states $\x$ satisfying $f(\x) = 1$ and $\ket{\psi_0}$ is a normalized superposition of ``bad'' states $\x$ satisfying $f(\x) = 0$ (again, for simplicity we ignore extra junk bits). If we measure $A \ket{\0}$, we would measure a ``good'' $\x$ with probability $\sin^2(\theta)$.
The goal of amplitude estimation is to approximate the amplitude $\sin(\theta)$. Let us assume for convenience that there are $t>0$ good states and $n-t > 0$ bad states, and that $0<\sin(\theta)<\pi/2$.

We note that the quantum search iterate $Q = -A U_0 A^{-1} U_f$ has eigenvalues $\ket{\psi_+} = \frac{1}{\sqrt{2}} (\ket{\psi_0} + i \ket{\psi_1})$ and $\ket{\psi_-} = \frac{1}{\sqrt{2}}(\ket{\psi_0} - i \ket{\psi_1})$ with respective eigenvalues $e^{i 2 \theta}$ and $e^{-i2 \theta}$. It also has $2^n-2$ other eigenvectors; $t-1$ of them have eigenvalue $+1$ and are the states orthogonal to $\ket{\psi_1}$ that have support on the states $\ket{\x}$ where $f(\x) = 1$, and $n-t-1$ of them have eigenvalue $-1$  and are the states orthogonal to $\ket{\psi_0}$ that have support on the states $\ket{\x}$ where $f(\x) = 0$.
It is important to note that $A\ket{\psi} = \frac{e^{i \theta}}{\sqrt{2}} \ket{\psi_+} + \frac{e^{-i\theta}}{\sqrt{2}} \ket{\psi_-}$  has its full support on the two dimensional subspace spanned by $\ket{\psi_+}$ and $\ket{\psi_-}$.

 It is worth noting that the quantum search iterate $Q = -A U_0 A^{-1} U_f$ can also be thought of as two reflections \[ -U_{A\ket{\0}} U_f = (2 A\ket{\0} \bra{\0}A^{\dagger}-I)(I - 2 \sum_{\x | f(\x) = 1} \ket{\x}\bra{\x}) ,\]
 one which flags the ``good'' subspace with a $-1$ phase shift, and then one that flags the subspace orthogonal to $A\ket{\0}$ with a $-1$ phase shift. In the two dimensional subspace spanned by $A \ket{0}$ and $U_f A \ket{0}$, these two reflections correspond to a rotation by angle $2 \theta$. Thus it should not be surprising to find eigenvalues $e^{\pm 2 \pi i \theta}$ for states in this two dimensional subspace.
 (In the section on quantum walks, we'll consider a more general situation where we have an operator $Q$ with many non-trivial eigenspaces and eigenvalues.)

\begin{figure}[htp]
{\large{$$\Qcircuit @C=1.5em @R=.3em @!R  {
& \multigate{4}{QFT_M} & \multimeasure{4}{\!} & \multigate{4}{QFT_M^{-1}} & \measuretab{y_1}  \\
     & \ghost{QFT_M} & \ghost{\!} & \ghost{QFT_M^{-1}} & \measuretab{y_2}  \\
\lstick{\ket{0}^{\otimes m}}& \ghost{QFT_M} & \ghost{\!} & \ghost{QFT_M^{-1}} & \measuretab{y_3}&\lstick{y}  \\
     \rstick{\vdots}\\
     & \ghost{QFT_M} & \ghost{\!} \qwx[2]& \ghost{QFT_M^{-1}} & \measuretab{y_m} \\
     \\
& \multigate{4}{A} & \multigate{4}{Q^x} & \qw & \qw \\
     & \ghost{A} & \ghost{G^x} & \qw & \qw  \\
\lstick{\ket{0}^{\otimes n}}& \ghost{A} & \ghost{Q^x} & \qw & \qw  \\
     \rstick{\vdots}\\
     & \ghost{A} & \ghost{Q^x}& \qw &\qw\\
}$$}}
\caption{The above circuit estimates the amplitude with which the unitary operator $A$ maps $\ket{\0}$ to the subspace of solutions to $f(\x)=1$. One uses $m \in \Omega(\log (1/\epsilon))$ qubits in the top register, and prepares it in a uniform superposition of the strings $y$ representing the integers $0,1, 2, \ldots, 2^{m}-1$ (one can in fact optimize the amplitudes of each $y$ to achieve a slightly better estimate \cite{DDEMM07}). The controlled-$Q^y$ circuit applies $Q^y$ to the bottom register when the value $y$ is encoded in the top qubits. If $A \ket{\0} = \sin(\theta) \ket{\psi_1} + \cos(\theta) \ket{\psi_0}$, where $\ket{\psi_1}$ is a superposition of solutions $\x$ to $f(\x)=1$, and $\ket{\psi_0}$ is a superposition of values $\x$ with $f(\x)=0$, then the value $y = y_1 y_2 \ldots y_m$ measured in the top register corresponds to the phase estimate $2 \pi y/2^m$ which is likely to be within $\frac{2 \pi}{2^m}$ (modulo $2 \pi$) of either $2\theta$ or $-2\theta$. Thus the value of $\sin^2{ \pi y/2^m}$ is likely to satisfy $|\sin^2{ \pi y/2^m}-\sin^2{\theta}| \in O(\epsilon)$.}
\label{fig:amplitude-estimation}
\end{figure}
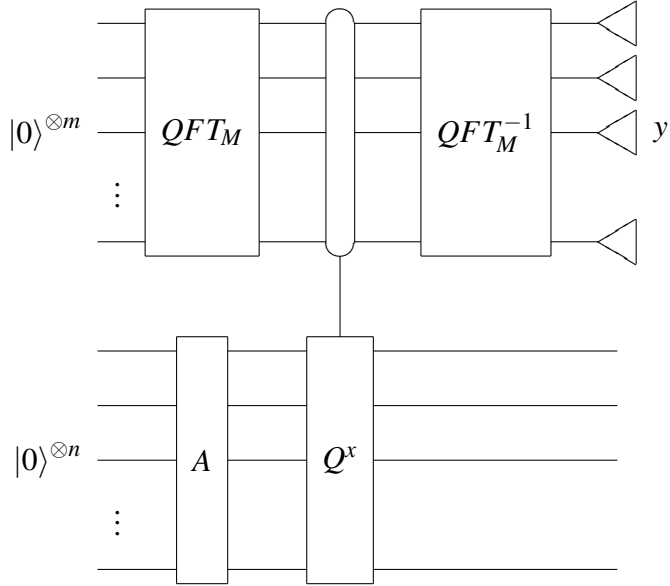

Thus one can approximate $\sin(\theta)$ by estimating the eigenvalue of either $\ket{\psi_+}$ or $\ket{\psi_-}$.
Performing the standard eigenvalue estimation algorithm on $Q$ with input $A\ket{\0}$ (as illustrated in Figure \ref{fig:amplitude-estimation}) gives a quadratic speed-up for estimating $\sin^2(\theta)$ versus simply repeatedly measuring $A\ket{\0}$ and counting the frequency of $1$s.
In particular, we can obtain an estimate $\tilde{\theta}$ of $\theta$ such that $|\tilde{\theta} - \theta| < \epsilon$ with high (constant) probability using $O(1/\epsilon)$ repetitions of $Q$, and thus $O(1/\epsilon)$ repetitions of $U_f$, $A$ and $A^{-1}$. For fixed $p$, this implies that $\tilde{p} = \sin^2(\tilde{\theta})$ satisfies $|p - \tilde{p}| \in O(\epsilon)$ with high probability. Classically sampling requires $O(1/\epsilon^2)$ samples for the same precision.
One application is speeding up the efficiency of the ``Hadamard'' test mentioned in section \ref{sec:hadamard-test}

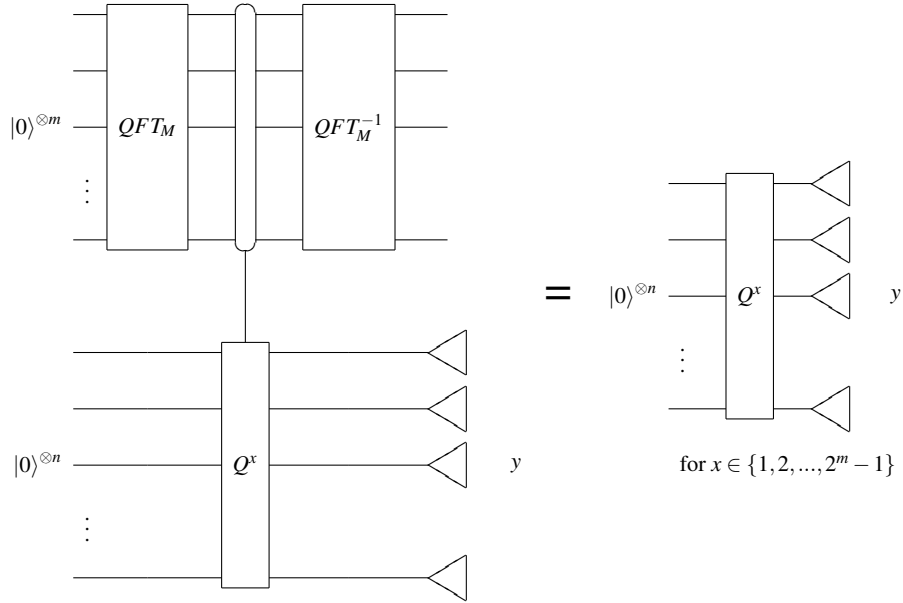
\begin{figure}[htp]
{{\footnotesize $$\Qcircuit @C=1.5em @R=.5em @!R  {
& \multigate{4}{QFT_M} & \multimeasure{4}{\!} & \multigate{4}{QFT_M^{-1}} &\qw  \\
     & \ghost{QFT_M} & \ghost{\!} & \ghost{QFT_M^{-1}}&\qw   \\
\lstick{\ket{0}^{\otimes m}}& \ghost{QFT_M} & \ghost{\!} & \ghost{QFT_M^{-1}} &\qw \\
     \rstick{\vdots}\\
     & \ghost{QFT_M} & \ghost{\!} \qwx[2]& \ghost{QFT_M^{-1}}&\qw  \\
     &&&&&&\rstick{\text{{\LARGE =}}}\\
& \qw & \multigate{4}{Q^x} & \qw &\measuretab{y_1}\\
     & \qw & \ghost{G^x} & \qw&\measuretab{y_2} \\
\lstick{\ket{0}^{\otimes n}}& \qw & \ghost{Q^x} & \qw&\measuretab{y_3}&\rstick{y}\\
     \rstick{\vdots}\\
     & \qw & \ghost{Q^x}& \qw&\measuretab{y_n}}
     \qquad\qquad\qquad
     \Qcircuit @C=1.5em @R=.5em @!R @!C {
     \\
     \\
     \\
 & \multigate{4}{Q^x} & \measuretab{y_1}\\
     & \ghost{G^x} & \measuretab{y_2}  \\
\lstick{\ket{0}^{\otimes n}}& \ghost{Q^x} & \measuretab{y_3}&\lstick{y}   \\
     \rstick{\vdots}\\
     & \ghost{Q^x}& \measuretab{y_n}\\
    \rstick{\text{for }x\in\{1,2,...,2^m-1\}}}
     $$}}
\caption{The amplitude estimation circuit can also be used for searching for a solution to $f(\x)=1$, as illustrated in the figure on the left. The second register starts off in the state $A \ket{\0}$, and after a sufficiently precise eigenvalue estimation of the quantum search iterate $Q$, the second register is approximately projected in the eigenbasis. The idea projection would yield the mixed state $\frac{1}{2} \ket{\psi_+}\bra{\psi_+} + \frac{1}{2} \ket{\psi_-}\bra{\psi_-} =  \frac{1}{2} \ket{\psi_1}\bra{\psi_1} + \frac{1}{2} \ket{\psi_0}\bra{\psi_0}$, and thus yields a solution to $f(\x)=1$ with probability approaching $\frac{1}{2}$ as $m$ gets large (the probability is in $\Omega(1)$ once $m \in \Omega(1/\sqrt{p})$).  The same algorithm can in fact be achieved by discarding the top register and instead randomly picking a value $y \in \{0,1, \ldots, 2^m-1\}$ and applying $Q^y$ to $A\ket{0}$, as illustrated on the right.}
\label{fig:quantum-searching}
\end{figure}

Another interesting observation is that increasingly precise eigenvalue estimation of $Q$ on input $A\ket{\0}$ leaves the eigenvector register in a state that gets closer and closer to the mixture $\frac{1}{2} \ket{\psi_+}\bra{\psi_+} + \frac{1}{2} \ket{\psi_-} \bra{\psi_-}$ which equals $\frac{1}{2} \ket{\psi_1}\bra{\psi_1} + \frac{1}{2} \ket{\psi_0} \bra{\psi_0}$. Thus eigenvalue estimation will leave the eigenvector register in a state that contains a solution to $f(\x)=1$ with probability approaching $\frac{1}{2}$.
One can in fact using this entire algorithm as subroutine in another quantum search algorithm, and obtain an algorithm with success probability approaching $1$ \cite{Mos01, KLM07}, and the convergence rate can be improved further \cite{TGP06}.
Another important observation is that for the purpose of searching, the eigenvalue estimate register is never used except in order to determine a random number of times in which to apply $Q$ to the second register. This in fact gives the quantum searching algorithm of \cite{BBHT98}. In other words, applying $Q$ a random number of times decoheres or approximately projects the eigenvector register in the eigenbasis, which gives a solution with high probability.  The method of approximating projections by using randomized evolutions was recently refined, generalized, and applied in \cite{BKS09}.

Quantum searching as discussed in this section has been further generalized in the quantum walk paradigm, which is the topic of the next section.

\section[1]{Quantum Walks} \label{sec:quantum-walks}
Random walks are a common tool throughout classical computer science.  Their applications include the simulation of biological, physical and social systems, as well as probabilistic algorithms such as Monte Carlo methods and the Metropolis algorithm.  A classical random walk is described by a $n\times n$ matrix $P$, where the entry $P_{u,v}$ is the probability of a transition from a vertex $v$ to an adjacent vertex $u$ in a graph $G$.  In order to preserve normalization, we require that $P$ is {\em stochastic}--- that is, the entries in each column must sum to 1.  We denote the initial probability distribution on the vertices of $G$ by the column vector $q$.  After $n$ steps of the random walk, the distribution is given by $P^nq$.

Quantum walks were developed in analogy to classical random walks, but it was not initially obvious how to do this.  Most generally, a quantum walk could begin in an initial state $\rho_0=\pr{\psi_0}$ and evolve according to any completely positive map $\mathcal{E}$ such that, after $t$ time steps, the system is in state $\rho(t)=\mathcal{E}^t(\rho_0)$. Such a quantum walk is simultaneously using classical randomness and quantum superposition. We can focus on the power of quantum mechanics by restricting to unitary walk operations, which maintain the system in a coherent quantum state.   So, the state at time $t$ can be described by $\ket{\psi(t)}=U^t\ket{\psi_0}$, for some unitary operator $U$.  However, it is not initially obvious how to define such a unitary operation.  A natural idea is to define the state space $A$ with basis $\{\ket{v}:v\in V(G)\}$ and walk operator $\tilde{P}$ defined by $\tilde{P}_{u,v}=\sqrt{P_{u,v}}$.  However, this will not generally yield a unitary operator $\tilde P$, and a more complex approach is required.  Some of the earliest formulations of unitary quantum walks appear in papers by Nayak and Vishwanath \cite{arXiv:quant-ph/0010117}, Ambainis et al. \cite{380757}, Kempe \cite{Kempe:2003sj}, and Aharonov et al. \cite{380758}.  These early works focused mainly on quantum walks on the line or a cycle.  In order to allow unitary evolution, the state space consisted of the vertex set of the graph, along with an extra ``coin register."  The state of the coin register is a superposition of $\ket{\text{\sc Left}}$ and $\ket{\text{\sc Right}}$.  The walk then proceeds by alternately taking a step in the direction dictated by the coin register and applying a unitary ``coin tossing operator" to the coin register.  The coin tossing operator is often chosen to be the Hadamard gate.  It was shown in \cite{380758,Ambainis:2003hl,380757,Kempe:2003sj,arXiv:quant-ph/0010117} that the mixing and propagation behaviour of these quantum walks was significantly different from their classical counterparts.  These early constructions developed into the more general concept of a discrete time quantum walk, which will be defined in detail.

We will describe two methods for defining a unitary walk operator.  In a {\em discrete time} quantum walk, the state space has basis vectors $\{\ket{u}\otimes\ket{v}:u,v\in V\}$.  Roughly speaking, the walk operator alternately takes steps in the first and second registers.  This is often described as a walk on the edges of the graph.  In a {\em continuous time} quantum walk, we will restrict our attention to symmetric transition matrices $P$.  We take $P$ to be the Hamiltonian for our system.  Applying Schr\"{o}dinger's equation, this will define continuous time unitary evolution in the state space spanned by $\{\ket{u}:u\in V\}$.  Interestingly, these two types of walk are not known to be equivalent.  We will give an overview of both types of walk as well as some of the algorithms that apply them.

\subsection{Discrete Time Quantum Walks}

Let $P$ be a stochastic matrix describing a classical random walk on a graph $G$.  We would like the quantum walk to respect the structure of the graph $G$, and take into account the transition probabilities $P_{u,v}$.  The quantum walk should be governed by a unitary operation, and is therefore reversible.  However, a classical random walk is not, in general, a reversible process.  Therefore, the quantum walk will necessarily behave differently than the classical walk.  While the state space of the classical walk is $V$, the state quantum walk takes place in the space spanned by $\{\ket{u,v}:\ths u,v\in V\}$.  We can think of the first register as the current location of the walk, and the second register as a record of the previous location.  To facilitate the quantum walk from a state $\ket{u,v}$, we first mix the second register over the neighbours of $u$, and then swap the two registers.  The method by which we mix over the neighbours of $u$ must be chosen carefully to ensure that it is unitary.  To describe this formally, we define the following states for each $u\in V$:
\bea \ket{\psi_u}=\ket{u}\otimes\sum_{v\in V}\sqrt{P_{vu}}\ket{v}\\\ket{\psi_u^*}=\sum_{v\in V}\sqrt{P_{vu}}\ket{v}\otimes\ket{u}.\ts\eea
Furthermore, define the projections onto the space spanned by these states:
\bea \Pi=\sum_{u\in V}\density{\psi_u}\\ \Pi^*=\sum_{u\in V}\density{\psi_u^*}.\ts\eea
In order to mix the second register, we perform the reflection $(2\Pi-I)$.  Letting $S$ denote the swap operation, this process can be written as \be S(2\Pi-I).\ts\ee
It turns out that we will get a more elegant expression for a single step of the quantum walk if we define the walk operator $W$ to be two iterations of this process:
\begin{flalign} W=&S(2\Pi-I)S(2\Pi-I)\\=&(2(S\Pi S)-I)(2\Pi-I)\\=&(2\Pi^*-I)(2\Pi-I).\ts\end{flalign}
So, the walk operator $W$ is equivalent to performing two reflections.

Many of the useful properties of quantum walks can be understood in terms of the spectrum of the operator $W$.  First, we define $D$, the $n\times n$ matrix with entries $D_{u,v}=\sqrt{P_{u,v}P_{v,u}}$.  This is called the {\em discriminant matrix}, and has eigenvalues in the interval $[0,1]$.  In the theorem that follows, the eigenvalues of $D$ that lie in the interval $(0,1)$ will be expressed as $\cos(\theta_1),...,\cos(\theta_k)$.  Let $\ket{\theta_1},...,\ket{\theta_k}$ be the corresponding eigenvectors of $D$.   Now, define the subspaces
\bea\label{spaceA} A=span\left\{\Ket{\psi_u}\right\}\\
\label{spaceB}B=span\left\{\Ket{\psi_u^*}\right\}.\ts\eea
Finally, define the operator \be Q=\sum_{v\in V}\ketbra{\psi_v}{v}\ee and \be\ket{\phi_j}=Q\ket{\theta_j}.\ts\ee  We can now state the following spectral theorem for quantum walks:
\begin{theorem}[Szegedy, \cite{1033158}]\label{spectral} The eigenvalues of $W$ acting on the space $A+B$ can be described as follows:
\begin{enumerate}
\item The eigenvalues of $W$ with non-zero imaginary part are $e^{\pm2i\theta_1},e^{\pm2i\theta_2},...,e^{\pm2i\theta_k}$ where $\cos(\theta_1),...,\cos(\theta_k)$ are the eigenvalues of $D$ in the interval $(0,1)$.  The corresponding (un-normalized) eigenvectors of $W(P)$ can be written as $\ket{\phi_j}-e^{\pm2i\theta_j}S\ket{\phi_j}$ for $j=1,...,k$.
\item $A\cap B$ and $A^\perp\cap B^\perp$ span the $+1$ eigenspace of $W$.  There is a direct correspondence between this space and the $+1$ eigenspace of $D$.  In particular, the $+1$ eigenspace of $W$ has the same degeneracy as the $+1$ eigenspace of $D$.
\item $A\cap B^\perp$ and $A^\perp\cap B$ span the $-1$ eigenspace of $W$.
\end{enumerate}
\end{theorem}
We say that $P$ is {\em symmetric} if $P^T=P$ and {\em ergodic} if it is aperiodic.  Note that if $P$ is symmetric, then the eigenvalues of $D$ are just the absolute values of the eigenvalues of $P$.  It is well-known that if $P$ is ergodic, then it has exactly one stationary distribution (i.e. a unique $+1$ eigenvalue).  Combining this fact with theorem (\ref{spectral}) gives us the following corollary:
\begin{corollary} If $P$ is ergodic and symmetric, then the corresponding walk operator $W$ has unique $+1$ eigenvector in $Span(A,B)$: \be\label{psi}\ket{\psi}=\frac{1}{\sqrt{n}}\sum_{v\in V}\ket{\psi_v}=\frac{1}{\sqrt{n}}\sum_{v\in V}\ket{\psi_v^*}\ee Moreover, if we measure the first register of $\ket{\psi}$, we get a state corresponding to vertex $u$ with probability \be Pr(u)=\frac{1}{n}\sum_{v\in V}P_{u,v}=\frac{1}{n}.\ts\ee This is the uniform distribution, which is the unique stationary distribution for the classical random walk.\end{corollary}

\subsubsection{The Phase Gap and the Detection Problem}
In this section, we will give an example of a quadratic speedup for the problem of detecting whether there are any ``marked" vertices in the graph $G$.  First, we define the following:
\begin{definition}The {\em phase gap} of a quantum walk is defined as the smallest postive value $2\theta$ such that $e^{\pm2i\theta}$ are eigenvalues of the quantum walk operator.  It is denoted by $\Delta(P)$.\end{definition}
\begin{definition} Let $M\subseteq V$ be a set of marked vertices.  In the {\em detection problem}, we are asked to decide whether $M$ is empty.\end{definition}
In this problem, we assume that $P$ is symmetric and ergodic.  We define the following modified walk $P\pri$:
\be P\pri_{uv}=\begin{cases}P_{uv}&v\notin{M}\\0&u\neq v, v\in M\\1&u=v, v\in M.\ts\end{cases}\ee
This walk resembles $P$, except that it acts as the identity on the set $M$.  That is, if the walk reaches a marked vertex, it stays there.  Let $P_{M}$ denote the operator $P\pri$ restricted to $V\setminus M$.  Then, arranging the rows and columns of $P\pri$, we can write \be P\pri=\begin{pmatrix}P_M&0\\P^{\prime\prime}&I\end{pmatrix}.\ts\ee By {\bf Theorem \ref{spectral}},  if $M=\emptyset$, then $P_M=P\pri=P$ and $\vnorm{P_M}=1$.  Otherwise, we have the strict inequality $\vnorm{P_M}<1$.  The following theorem bounds $\vnorm{P_M}$ away from 1:
\begin{theorem}\label{normbound} If $(1-\delta)$ is the absolute value of the eigenvalue of $P$ with second largest magnitude, and $\norm{M}\geq\epsilon\norm{V}$, then $\vnorm{P_M}\leq 1-\frac{\delta\epsilon}{2}$.\end{theorem}
We will now show that the detection problem can be solved using eigenvalue estimation.  {\bf Theorem \ref{normbound}} will allow us to bound the running time of this method.  First, we describe the discriminant matrix for $P\pri$:
\be D(P\pri)_{uv}=\begin{cases}P_{uv}&u,v\notin{M}\\1&u=v, v\in M\\0&\text{otherwise}.\ts\end{cases}\ee
Now, beginning with the state \be\ket{\psi}=\frac{1}{\sqrt{n}}\sum_{v\in V}\ket{\psi_v}=\frac{1}{\sqrt{n}}\sum_{v\in V}\sqrt{P_{v,u}}\ket{u}\ket{v},\ee
we measure whether or not we have a marked vertex; if so, we are done.  Otherwise, we have the state
\be\ket{\psi_M}=\frac{1}{\sqrt{\norm{V\setminus M}}}\sum_{u,v\in V\setminus M}\sqrt{P_{vu}}\ket{u}\ket{v}.\ts\ee
If $M=\emptyset$, then this is the state $\ket{\psi}$ defined in (\ref{psi}), and is the $+1$ eigenvector of $W(P)$.  Otherwise, by {\bf Theorem \ref{spectral}}, this state lies entirely in the space spanned by eigenvectors with values of the form $e^{\pm2i\theta_j}$, where $\theta_j$ is an eigenvalue of $P_M$.  Applying {\bf Theorem \ref{normbound}}, we know that \be\theta\geq \cos^{-1}(1-\frac{\delta\epsilon}{2})\geq\sqrt{\frac{\delta\epsilon}{2}}.\ts\ee So, the task of distinguishing between $M$ being empty or non-empty is equivalent to that of distinguishing between a phase parameter of $0$ and a phase parameter of at least $\sqrt{\frac{\delta\epsilon}{2}}$.  Therefore, applying phase estimation to $W(P\pri)$ on state $\ket{\psi_M}$ with precision $O(\sqrt{\delta\epsilon})$ will decide whether $M$ is empty with constant probability.  This requires time $O(\frac{1}{\sqrt{\delta\epsilon}})$.

By considering the modified walk operator $P\pri$, it can be shown that the detection problem requires $O(\frac{1}{\delta\epsilon})$ time in the classical setting.  Therefore the quantum algorithm provides a quadratic speedup over the classical one for the detection problem.

\subsubsection{Quantum Hitting Time}\label{QHT}

Classically, the first hitting time is denoted $H(\rho,M)$.  For a walk defined by $P$, starting from the probability distribution $\rho$ on $V$, $H(\rho,M)$ is the smallest value $n$ such that the walk reaches a marked vertex $v\in M$ at some time $t\in\{0,...,n\}$ with constant probability.  This idea is captured by applying the modified operator $P\pri$ and some $n$ times, and then considering the probability that the walk is in some marked state $v\in M$.  Let $\rho_M$ be any initial distribution restricted to the vertices $V\setminus M$.  Then, at time $t$, the probability that the walk is in an unmarked state is $\vnorm{P_M^t\rho_M}_1$, where $\vnorm{\cdot}_1$ denotes the $L_1$ norm.  Assuming that $M$ is non-empty, we can see that $\vnorm{P_M}<1$.  So, as $t\rightarrow\infty$, we have $\vnorm{P_M^t\rho_M}_1\rightarrow0$.  So, as $t\rightarrow\infty$, the walk defined by $P\pri$ is in a marked state with probability 1.  As a result, if we begin in the uniform distribution  $\pi$ on $V$, and run the walk for some time $t$, we will ``skew" the distribution towards $M$, and thus away from the unmarked vertices.  So, we define the classical hitting time to be the minimum $t$ such that \be\vnorm{P_M^t\rho_M}_1<\epsilon\ee for any constant $\epsilon$ of our choosing.  Since the quantum walk is governed by a unitary operator, it doesn't converge to a particular distribution the way that the classical walk does.  We cannot simply wait an adequate number of time steps and then measure the state of the walk; the walk might have already been in a state with high overlap on the marked vertices and then evolved away from this state!  Quantum searching has the same problem when the number of solutions is unknown. We can get around this in a similar way by considering an expected value over a sufficiently long period of time.  This will form the basis of our definition of hitting time.  Define \be\label{pi}\ket\pi=\frac{1}{\sqrt{\size{V\setminus M}}}\sum_{v\in V\setminus M}\ket{\psi_v}.\ts\ee Then, if $M$ is empty, $\ket\pi$ is a $+1$ eigenvector of $W$, and $W^t\ket\pi=\ket\pi$ for all $t$.  However, if $M$ is non-empty, then the spectral theorem tells us that $\ket\pi$ lies in the space spanned by eigenvectors with eigenvalues $e^{\pm2i\theta_j}$ for non-zero $\theta_j$.  As a result, it can be shown that, for some values of $t$, the state $W^t\ket\pi$ is ``far" from the initial distribution $\ket\pi$.  We define the quantum hitting in the same way as Szegedy \cite{1033158}.  The hitting time $H_Q(W)$ as the minimum value $T$ such that \be\frac{1}{T+1}\sum_{t=0}^T\vnorm{W^t\ket{\pi}-\ket{\pi}}^2\geq1-\frac{\size{M}}{\size{V}}.\ee  This leads us to Szegedy's hitting time theorem \cite{1033158}:
\begin{theorem}\label{QHTheorem} The quantum hitting time $H_Q(W)$ is $$O\left(\frac{1}{\sqrt{1-\vnorm{P_M}}}\right)$$.  \end{theorem}
\begin{corollary}  \label{deltaepsilon}Applying {\bf Theorem \ref{normbound}}, if the second largest eigenvalue of $P$ has magnitude $(1-\delta)$ and $\size{M}/\size{V}\geq\epsilon$, then $H_Q(W)\in O(\frac{1}{\sqrt{\delta\epsilon}})$\end{corollary}
Notice that this corresponds to the running time for the algorithm for the detection problem, as described above.  Similarly, the classical hitting time is in $O(\frac{1}{\delta\epsilon})$, corresponding to the best classical algorithm for the detection problem.

It is also worth noting that, if there no marked elements, then the system remains in the state $\ket{\pi}$, and the algorithm never ``hits."  This gives us an alternative way to approach the detection problem.  We run the algorithm for  a randomly selected number of steps $t\in\{0,...,T\}$ with $T$ of size $O(\sqrt{1/\delta\epsilon})$, and then measure whether the system is still in the state $\ket\pi$; if there are any marked elements, then we can expect to find some other state with constant probability.

\subsubsection{The Element Distinctness Problem}
In the element distinctness problem, we are given a black box that computes the function
\be f:\{1,...,n\}\rightarrow S\ee
and we are asked to determine whether there exist $x, y\in\{1,...,n\}$ with $x\neq y$ and $f(x)=f(y)$.  We would like to minimize the number of queries made to the black box.  There is a lower bound of $\Omega(n^{2/3})$ on the number of queries, due indirectly to Aaronson and Shi \cite{1008735}.  The algorithm of Ambainis \cite{1366221} proves that this bound is tight.  The algorithm uses a quantum walk on the Johnson graph $J(n,p)$ has vertex set consisting of all subsets of $\{1,...,n\}$ of size $p$.  Let $S_1$ and $S_2$ be $p$-subsets of $\{1,...,n\}$.  Then, $S_1$ is adjacent to $S_2$ if and only if $\size{S_1\cap S_2}=p-1$.  The Johnson graph $J(n,p)$ therefore has $\binom{n}{p}$ vertices, each with degree $p(n-p)$.

The state corresponding to a vertex $S$ of the Johnson graph will not only record which subset $\{s_1,...,s_p\}\subseteq\{1,...,n\}$ it represents, but the function values of those elements.  That is,
\be\ket{S}=\ket{s_1,s_2...,s_p;f(s_1),f(s_2),...,f(s_p)}.\ts\ee
Setting up such a state requires $p$ queries to the black box.

The walk then proceeds for $t$ iteration, where $t$ is chosen from the uniform distribution on $\{0,...,T\}$ and $T\in O(1/\sqrt{\delta\epsilon})$.  Each iteration has two parts to it.  First, we need to check if there are distinct $s_i, s_j$ with $f(s_i)=f(s_j)$---that is, whether the vertex $S$ is marked.  This requires no calls to the black box, since the function values are stored in the state itself.  Second, if the state is unmarked, we need to take a step of the walk.  This involves replacing, say, $s_i$ with $s_i\pri$, requiring one query to erase $f(s_i)$ and another to insert the value $f(s_i\pri)$.  So, each iteration requires a total of 2 queries.

We will now bound $\epsilon$ and $\delta$.  If only one pair $x,y$ exists with $f(x)=f(y)$, then there are $\binom{n-2}{p-2}$ marked vertices.  This tells us that, if there are any such pairs at all, epsilon is $\Omega({p^2}/{n^2})$.  Johnson graphs are very well-understood in graph theory.  It is a well known result that the eigenvalues for the associated walk operator are given by:
\be\lambda_j=1-\frac{j(n+1-j)}{p(n-p)}\ee
for $0\leq j\leq p$.  For a proof, see \cite{Brouwer:1989bf}.  This give us $\delta=\frac{n}{p(n-p)}$.  Putting this together, we find that $\sqrt{1/\delta\epsilon}$ is $O(\frac{n}{\sqrt{p}})$.  So, the number of queries required is $O(p+\frac{n}{\sqrt{p}})$.  In order to minimize this quantity, we choose $m$ to be of size $\Theta(n^{2/3})$.  The query complexity of this algorithm is $O(n^{2/3})$, matching the lower bound of Aaronson and Shi \cite{1008735}.

\subsubsection{Unstructured Search as a Discrete Time Quantum Walk}
\begin{figure}[htp]
\sidecaption
\label{fig:reflection}
\includegraphics[scale=0.9]{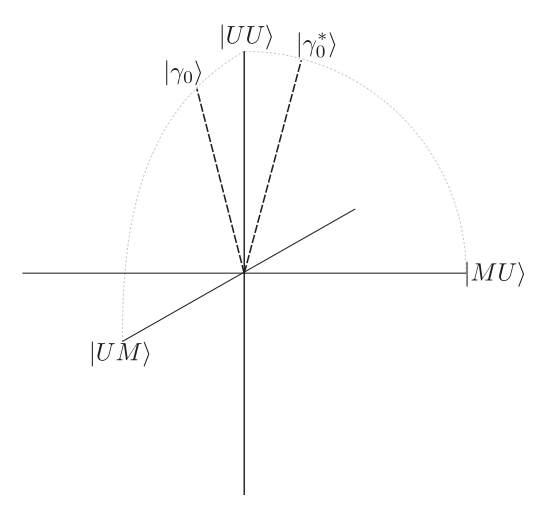}
\caption{The 3 dimensional space $\Gamma$.  The walk operator acts on this space by alternately performing reflections in the vector $\ket{\gamma_0}$ and $\ket{\gamma_0^*}$.}
\end{figure}

We will now consider unstructured search in terms of quantum walks.  For unstructured search, we are required to identify a marked element from a set of size $n$.  Let $M$ denote the set of marked elements, and $U$ denote the set of unmarked elements.  Furthermore, let $m=\size{M}$ and $q=\size{U}$.  We assume that the number of marked elements, $m$, is very small in relation to the total number of elements $n$.  If this were not the case, we could efficiently find a marked element with high probability by simply checking a randomly chosen set of vertices.  Since the set lacks any structure or ordering, the corresponding walk takes place on the complete graph on $n$ vertices.
Let us define the following three states:
\bea\ket{UU}=\frac{1}{q}\sum_{u,v\in U}\ket{u,v}\\
\ket{UM}=\frac{1}{\sqrt{nq}}\sum_{\substack{u\in U\\v\in M}}\ket{u,v}\\
\ket{MU}=\frac{1}{\sqrt{nq}}\sum_{\substack{u\in M\\v\in U}}\ket{u,v}.\ts\eea
Noting that $\{\ket{UU}, \ket{UM}, \ket{MU}\}$ is an orthonormal set, we will consider the action of the walk operator on the three dimensional space \be \Gamma=span\{\ket{UU}, \ket{UM}, \ket{MU}\}.\ts\ee
In order to do this, we will express the spaces $A$ and $B$, defined in (\ref{spaceA}-\ref{spaceB}) in terms of a different basis.  First we label the unmarked vertices:
\be U=\{u_0,...,u_{q-1}\}.\ts\ee
Then, we define
\be \ket{\gamma_k}=\frac{1}{\sqrt{q}}\sum_{j=0}^{q-1}e^{\frac{2\pi ijk}{q}}\ket{\psi_{u_j}}\ee
and
\be \ket{\gamma^*_k}=\frac{1}{\sqrt{q}}\sum_{j=0}^{q-1}e^{\frac{2\pi ijk}{q}}\ket{\psi^*_{u_j}}\ee
with $k$ ranging from $0$ to $q-1$.  Note that $\ket{\gamma_0}$ corresponds to the definition of $\ket{\pi}$ in (\ref{pi}).  We can then rewrite $A$ and $B$:
\bea A=span\{\ket{\gamma_k}\}\\
B=span\{\ket{\gamma^*_k}\}.\ts
\eea
Now, note that for $k\neq0$, the space $\Gamma$ is orthogonal to $\ket{\gamma_k}$ and $\ket{\gamma^*_k}$.  Furthermore,
\be \ket{\gamma_0}=\frac{1}{\sqrt{n}}\left(\sqrt{m}\ket{UM}+\sqrt{q}\ket{UU}\right)\ee
and
\be \ket{\gamma^*_0}=\frac{1}{\sqrt{n}}\left(\sqrt{m}\ket{MU}+\sqrt{q}\ket{UU}\right).\ts\ee
Therefore, the walk operator
\be W=(2\Pi^*-I)(2\Pi-I)\ee
when restricted to $\Gamma$ is simply
\be W\pri=\Bigl(2\pr{\gamma^*_0}-I\Bigr)\Bigl(2\pr{\gamma_0}-I\Bigr).\ts\ee
Figure \ref{fig:reflection} illustrates the space $\Gamma$ which contains the vectors $\ket{\gamma_0}$ and $\ket{\gamma_0^*}$.

At this point, it is interesting to compare this algorithm to Grover's search algorithm.  Define \be\ket{\rho_t}=W^t\ket{UU}.\ts\ee Now, define $\ket{\rho_t^1}$ and $\ket{\rho_t^2}$ to be the projection of $\ket{\rho_t}$ onto $span\{\ket{UU}, \ket{UM}\}$ and $span\{\ket{UU}, \ket{MU}\}$ respectively.  We can think of these as the ``shadow" cast by the vector $\ket{\rho_t}$ on two-dimensional planes within the space $\Gamma$.  Note that $\ket{\gamma_0}$ lies in the $span\{\ket{UU}, \ket{UM}\}$ and its projection onto $span\{\ket{UU}, \ket{MU}\}$ is $\sqrt{q/n}\ket{UU}$.  So, the walk operator acts on $\ket{\rho_t^1}$ by reflecting it around the vector $\ket{\gamma_0}$ and then around $\sqrt{q/n}\ket{UU}$.  This is very similar to Grover search, except for the fact that the walk operator in this case does not preserve the magnitude of $\ket{\rho_1^t}$.  So, with each application of the walk operator, $\ket{\rho_t^1}$ is rotated by $2\theta$, where $\theta=\tan^{-1}(\sqrt{m/q})$ is the angle between $\ket{UU}$ and $\ket{\gamma_0}$.  The case for $\ket{\rho_t^2}$ is exactly analogous, with the rotation taking place in the plane $span\{\ket{UU}, \ket{MU}\}$.  So, we can think of the quantum walk as Grover search taking place simultaneously in two intersecting planes.  It  should not come as a surprise, then, that we can achieve a quadratic speedup similar to that of Grover search.

We will now use a slightly modified definition of hitting time to show that the walk can indeed be used to find a marked vertex in $O(\sqrt{n})$ time.  Rather than using the hitting time as defined in section \ref{QHT}, we will replace the state $\ket{\pi}=\ket{\gamma_0}$ with $\ket{UU}$.  Note that we can create $\ket{UU}$ from $\ket{\pi}$ by simply measuring whether the second register contains a marked vertex.  This is only a small adjustment, since $\ket{\gamma_0}$ is close to $\ket{UU}$.  Furthermore, both lie in the space $\Gamma$, and the action of the walk operator is essentially identical for both starting states.  It should not be surprising to the reader that the results of section \ref{QHT} apply to this modified definition as well.

In this case, the operator $P$ is defined by:
\be P_{uv}=\begin{cases}0&\text{if }u=v\\ \frac{1}{n-1}&\text{if }u\neq v.\ts\end{cases}\ee
Let $v_0,...,v_{n-1}$ be a labeling of the vertices of $G$.  Let $x^0,...,x^{n-1}$ denote the eigenvectors of $P$, with $x^k_{v_j}$ denoting the amplitude on $v_j$ in $x^k$.  Then, the eigenvalues of $P$ are as follows:
\be x^k_{v_j}=\frac{1}{n}\cdot e^{\frac{2\pi ijk}{n}}.\ts\ee
Then, $x^0$ has eigenvalue 1, and is the stationary distribution.  All the other $x^k$ have eigenvalue $-1/(n-1)$, giving a spectral gap of $\delta=\frac{n-2}{n-1}$.  Applying corollary \ref{deltaepsilon}, this gives us quantum hitting time for the corresponding quantum walk operator $W$
\be H_Q(W)\leq\sqrt{\frac{(n-1)n}{n-2}}=O(\sqrt{n}).\ts\ee
So, we run the walk for some randomly selected time $t\in\{0,...,T-1\}$ with $T$ of size $O(\sqrt{n})$, then measure whether either the first or second register contains a marked vertex.  Applying theorem \ref{QHTheorem}, the probability that neither contains a marked vertex is
\begin{flalign*}\frac{1}{T}\sum_{t=0}^{T-1}\size{\braket{\rho_t}{UU}}^2\leq&\frac{1}{T}\sum_{t=0}^{T-1}\size{\braket{\rho_t}{UU}}\\
=&1-\frac{1}{2T}\sum_{t=0}^{T-1}\vnorm{\ket{\rho_t}-\ket{UU}}^2\\
\leq&1-\frac{\size{M}}{2\size{V}}
\end{flalign*}
Assuming that $\size{M}$ is small compared to $\size{V}$, we find a marked vertex in either the first or second register with high probability.  Repeating this procedure a constant number of times, our success probability can be forced arbitrarily close to 1.
%

\subsubsection{The MNRS Algorithm}
Magniez, Nayak, Roland and Santha \cite{1250874} developed an algorithm that generalizes the search algorithm described above to any graph.  A brief overview of this algorithm and others is also given in the survey paper by Santha \cite{Santha:2008ph}.  The MNRS algorithm employs similar principles to Grover's algorithm; we apply a reflection in $M$, the space of unmarked states, followed by a reflection in $\ket\pi$, a superposition of marked and unmarked states.  This facilitates a rotation through an angle related to the number of marked states.  In the general case considered by Magniez et al., $\ket\pi$ is the stationary distribution of the walk operator.  It turns out to be quite difficult to implement a reflection in $\ket\pi$ exactly.  Rather, the MNRS algorithm employs an approximate version of this reflection.  This algorithm requires $O\left(\frac{1}{\sqrt{\delta\epsilon}}\right)$ applications of the walk operator, where $\delta$ is the eigenvalue gap of the operator $P$, and $\epsilon$ is the proportion vertices that are marked.  In his survey paper, Santha \cite{Santha:2008ph} outlines some applications of the MNRS algorithm, including a version of the element distinctness problem where we are asked to find elements $x$ and $y$ such that $f(x)=f(y)$.

\subsection{Continuous Time Quantum Walks}
To define a classical continuous time random walk for a graph with no loops, we define a matrix similar to the adjacency matrix of $G$, called the {\em Laplacian}:
\be L_{uv}=\begin{cases}0&u\neq v, uv\notin E\\1&u\neq v, uv\in E\\-deg(v)&u=v.\ts\end{cases}\ee
Then, given a probability distribution $p(t)$ on the vertices of $G$, the walk is defined by
\be\ddt p(t)=Lp(t).\ts\ee Using the Laplacian rather than the adjacency matrix ensures that $p(t)$ remains normalized.  A continuous time quantum walk is defined in a similar way.  For simplicity, we will assume that the Laplacian is symmetric, although it is still possible to define the walk in the asymmetric case.  Then, since the Laplacian is Hermitian, we can simply take it to be the Hamiltonian of our system.  Letting $\ket{\rho(t)}$ be a normalized vector in $\mathbf{C}^V$, Schr\"odinger's equation gives us \be i\ddt\ket{\rho(t)}=L\ket{\rho(t)}.\ts\ee
Solving this equation, we get an explicit expression for $\ket{\rho(t)}$:
\be \ket{\rho(t)}=e^{-iLt}\ket{\rho(0)}.\ts\ee
Let $\{\ket{\lambda_1}.,,,\ket{\lambda_n}\}$ be the eigenvectors of $L$ with corresponding values $\{\lambda_1,...,\lambda_n\}$.  We can rewrite the expression for $\ket{\rho(t)}$ in terms of this basis as follows:
\be \ket{\rho(t)}=\sum_{j=1}^ne^{-i\lambda_jt}\braket{\lambda_j}{\rho_0}\ket{\lambda_j}.\ts\ee
Clearly, the behaviour of a continuous time quantum walk is very closely related to the eigenvectors and spectrum of the Laplacian.

Note that we are not required to take the Laplacian as the Hamiltonian for the walk.  We could take any Hermitian matrix we like, including the adjacency matrix, or the transition matrix of a (symmetric) Markov chain.

\subsubsection{A Continuous Time Walk for Unstructured Search}
For unstructured search, the walk takes place on a complete graph.  First, we define the following three states:
\begin{eqnarray}
\ket{V}=\frac{1}{\sqrt{\size{V}}}\sum_{v\in V}\ket{v}\\
\ket{M}=\frac{1}{\sqrt{\size{M}}}\sum_{v\in M}\ket{v}.
\end{eqnarray}
The Hamiltonian we will use is a slightly modified version of the Laplacian for the complete graph, with an extra ``marking term" added:
\be H=\density{V}+\density{M} .\ts\ee
It is convenient to consider the action of this Hamiltonian in terms of the vectors $\ket{M}$ and $\ket{M^\perp}$, where
\be \ket{M^\perp}=\frac{1}{\sqrt{\size{V\setminus M}}}\sum_{v\in V\setminus M}\ket{v}=\frac{1}{\sqrt{1-\braket{M}{V}^2}}(\ket{S}-\ket{M}\braket{M}{S}).\ts\ee
As outlined in \cite{Childs:im}, we let $\alpha=\braket{M}{S}$.  We can then re-write the Hamiltonian in terms of the basis $\{\ket{M},\ket{M^\perp}\}$:
\begin{flalign}
H=&\tbt{\alpha^2}{\alpha\sqrt{1-\alpha^2}}{\alpha\sqrt{1-\alpha^2}}{1-\alpha^2}+\tbt{1}{0}{0}{0}\\
=&\tbt{1}{0}{0}{1}+\alpha^2\tbt{1}{0}{0}{-1}+\alpha\sqrt{1-\alpha^2}\tbt{0}{1}{1}{0}\\
=&I+\alpha^2 \sigma_Z+\alpha\sqrt{1-\alpha^2}\sigma_X\\
=&I+\alpha\bigl(\alpha \sigma_Z+\sqrt{1-\alpha^2}\sigma_X\bigr)
\end{flalign}
where $\sigma_X$ and $\sigma_Z$ are the Pauli X and Z operators.  Note that the identity term in the sum simply introduces a global phase, and can be ignored.  Note that the operator
\be A=\alpha \sigma_Z+\sqrt{1-\alpha^2}\sigma_X\ee
has eigenvalues $\pm 1$, and is therefore Hermitian and unitary.  Therefore, we can write
\be e^{-iHt}=e^{-iA\alpha t}=\cos(\alpha t)I-i\sin(\alpha t)A.\ts\ee
Also note that
\begin{flalign}A\ket{S}=&\bigl(\alpha \sigma_Z+\sqrt{1-\alpha^2}\sigma_X\bigr)\bigl(\alpha\ket{M}+\sqrt{1-\alpha^2}\ket{M^\perp}\bigr)\\=&\ket{M}.\ts\end{flalign}
If we start with the state $\ket{S}$, we can now calculate the state of the system at time $t$:
\begin{flalign}
\ket{\rho(t)}=&\cos(\alpha t)\ket{S}-i\sin(\alpha t)A\ket{S}\\
=&\cos(\alpha t)\ket{S}-i\sin(\alpha t)\ket{M}.\ts
\end{flalign}
So, at time $t$, the probability of finding the system in a state corresponding to a marked vertex is
\begin{flalign}
\sum_{v\in M}\size{\braket{v}{\rho(t)}}^2=&\size{M}\left(\frac{\cos^2(\alpha t)}{\size{V}}+\frac{\sin^2(\alpha t)}{\size{M}}\right)\\
=&\alpha^2\cos^2(\alpha t)+\sin^2(\alpha t).\ts
\end{flalign}
At time $t=0$, this is the same as sampling from the uniform distribution, as we would expect.  However, at time $t=\pi/2\alpha=\pi/2\cdot\sqrt{N/M}$, we observe a marked vertex with probability 1.  Therefore, this search algorithm runs in time $O(\sqrt{N/M})$, which coincides with the running time of Grover's search algorithm and the related discrete time quantum walk algorithm.

\subsubsection{Mixing in Quantum Walks and the Glued Tree Traversal Problem}
\begin{figure}[htp]
\sidecaption
\includegraphics[scale=.6]{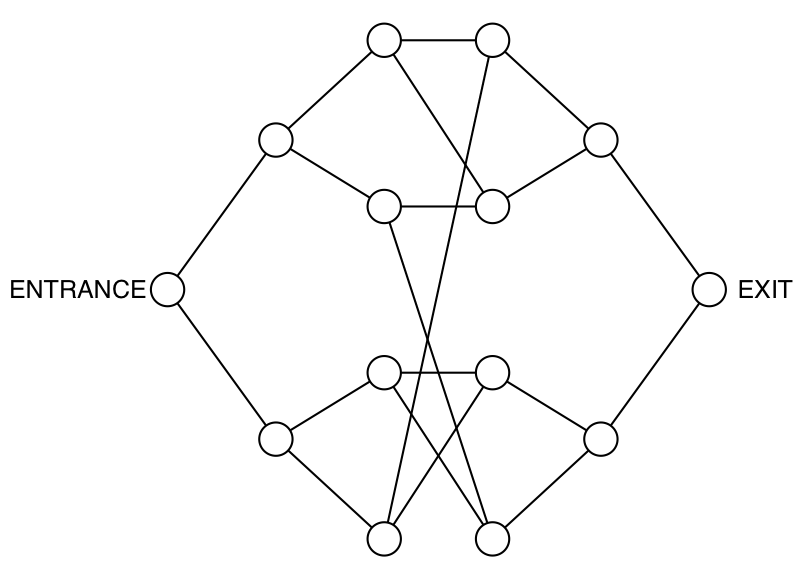}
\caption{A small glued tree graph.  }
\label{fig:gluedtrees}
\end{figure}

While continuous time quantum walks give us a generic quadratic speedup over their classical counterparts, there are some graphs on which the quantum walk gives exponential speedup.  One such example is the ``glued trees" of Childs et al. \cite{780552}.  In this example, the walk takes place on the following graph, obtained by taking two identical binary trees of depth $d$ and joining their leaves using a random cycle, as illustrated in Figure \ref{fig:gluedtrees}.
Beginning at the vertex {\sc entrance}, we would like to know how long we need to run the algorithm to reach {\sc exit}.  It is straightforward to show that classically, this will take time exponential in $d$, the depth of the tree.  Intuitively, this is because there are so many vertices and edges in the ``middle" of the graph, that there is small probability of a classical walk ``escaping" to the exit.  We will prove that a continuous time quantum walk can achieve an overlap of $\Omega(P(d))$ with the {\sc exit} vertex in time $O(\frac{1}{Q(d)})$, where $Q$ and $P$ are polynomials.

Let $V_s$ denote the set of vertices at depth $s$, so that $V_0=\{\text{{\sc entrance}}\}$ and $V_{2d+1}=\{\text{{\sc exit}}\}$.  Taking the adjacency matrix to be our Hamiltonian, the operator $U(t)=e^{-iHt}$ acts identically on the vertices in $V_s$ for any $s$.  Therefore, the states \be\ket{s}=\frac{1}{\sqrt{\size{V_s}}}\sum_{v\in V_s}\ket{v}\ee
form a convenient basis.  We can therefore think of the walk on $G$ as a walk on the states $\{\ket{s}:\ts0\leq s\leq 2d+1\}$.  We also note that, if $A$ is the adjacency matrix of $G$, 
\be A\ket{s}=\begin{cases}\sqrt{2}\ket{s+1}&s=0\\\sqrt{2}\ket{s-1}&s=2d+1\\\sqrt{2}\ket{s-1}+2\ket{s+1}&s=d\\\sqrt{2}\ket{s+1}+2\ket{s-1}&s=d+1\\\sqrt{2}\ket{s+1}+\sqrt{2}\ket{s-1}&\text{otherwise}.\ts\end{cases}\ee
Aside from the exceptions at {\sc entrance}, {\sc exit}, and the vertices in the center, this walk looks exactly like the walk on a line with uniform transition probabilities.

Continuous time classical random walks will eventually converge to a limiting distribution.  The time that it takes for the classical random walk to get `close' to this distribution is called the {\em mixing time}.  More formally, we can take some small constant $\gamma$, and take the mixing time to be the amount of time time it take to come within $\gamma$ of the stationary distribution, according to some metric.  In general, we express the mixing time in terms of $1/\gamma$ and $n$, the number of vertices in the graph.

Since quantum walks are governed by a unitary operator, we cannot expect the same convergent behaviour.  However, we can define the limiting behaviour of quantum walks by taking an average over time.  In order to do this, we define $\Pr(u, v, T)$.  If we select $t\in[0,T]$ uniformly at random and run the walk for time $t$, beginning at vertex $u$, then $\Pr(u, v, T)$ is the probability that we find the system in state $v$.  Formally, we can write this as
\be \Pr(u,v,T)=\frac{1}{T}\int_0^T\size{\bra{v}e^{-iAt}\ket{u}}^2dt\ee
where $A$ is the adjacency matrix of the graph.  Now, if we take $\{\ket{\lambda}\}$ to be the set of eigenvectors of $A$ with corresponding eigenvalues $\{\lambda\}$, then we can rewrite $\Pr(u,v,T)$ as follows:
\begin{flalign}
\Pr(u,v,T)=&\frac{1}{T}\int_0^T\size{\sum_\lambda e^{-i\lambda t}\braket{v}{\lambda}\braket{\lambda}{u}}^2dt\\
=&\frac{1}{T}\int_0^T\sum_{\lambda,\lambda\pri}\left(e^{-i(\lambda-\lambda\pri) t}\braket{v}{\lambda}\braket{\lambda}{u}\braket{v}{\lambda\pri}\braket{\lambda\pri}{u}\right)dt\\
=&\sum_{\lambda}\size{\braket{v}{\lambda}\braket{\lambda}{u}}^2\\&+\frac{1}{T}\sum_{\lambda\neq\lambda\pri}\braket{v}{\lambda}\braket{\lambda}{u}\braket{v}{\lambda\pri}\braket{\lambda\pri}{u}\int_0^Te^{-i(\lambda-\lambda\pri) t}dt\\
=&\sum_{\lambda}\size{\braket{v}{\lambda}\braket{\lambda}{u}}^2\\&+\sum_{\lambda\neq\lambda\pri}\braket{v}{\lambda}\braket{\lambda}{u}\braket{v}{\lambda\pri}\braket{\lambda\pri}{u}\frac{1-e^{-i(\lambda-\lambda\pri)T}}{i(\lambda-\lambda\pri)T}.\ts
\end{flalign}
In particular, we have
\be\label{prinf}\lim_{T\rightarrow\infty}\Pr(u,v,T)=\sum_{\lambda}\size{\braket{v}{\lambda}\braket{\lambda}{u}}^2.\ts\ee
We will denote this value by $\Pr(u,v,\infty)$.  This is the quantum analogue of the limiting distribution for a classical random walk.  We would now like to apply this to the specific case of the glued tree traversal problem.  First, we will lower bound $\Pr(\text{\sc entrance},\text{\sc exit},\infty)$.  We will then show that $\Pr(\text{\sc entrance},\text{\sc exit},T)$ approaches this value rapidly as we increase $T$, implying that we can traverse the glued tree structure efficiently using a quantum walk.

Define the reflection operator $\Theta$ by \be\Theta\ket{j}=\ket{2d-1-j}\ee This operator commutes with the adjacency matrix, and hence the walk operator because of the symmetry of the glued trees.  This implies that $\Theta$ can be diagonalized in the eigenbasis of the walk operator $e^{-iAt}$ for any $t$.  What is more, the eigenvalues of $\Theta$ are $\pm1$.  As a result, if $\ket\lambda$ is an eigenvalue of $e^{-iAt}$, then \be\braket{\lambda}{\text{\sc entrance}}=\pm\braket{\lambda}{\text{\sc exit}}.\ts\ee
We can apply this to (\ref{prinf}), yielding
\begin{flalign}\Pr(\text{\sc entrance},\text{\sc exit}, \infty)=&\sum_\lambda\size{\braket{\text{\sc entrance}}{\lambda}}^4\\
\geq&\frac{1}{2d+2}\sum_\lambda\size{\braket{\text{\sc entrance}}{\lambda}}^2\\
=&\frac{1}{2d+2}.\ts
\end{flalign}
Now, we need to determine how quickly $\Pr(\text{\sc entrance},\text{\sc exit}, T)$ approaches $\Pr(\text{\sc entrance},\text{\sc exit}, \infty)$ as we increase $T$:
\begin{flalign}
&\size{\Pr(\text{\sc entrance},\text{\sc exit}, T)-\Pr(\text{\sc entrance},\text{\sc exit}, \infty)}\\=&\size{\sum_{\lambda\neq\lambda\pri}\braket{\text{\sc exit}}{\lambda}\braket{\lambda}{\text{\sc entrance}}\braket{\text{\sc exit}}{\lambda\pri}\braket{\lambda\pri}{\text{\sc entrance}}\frac{1-e^{-i(\lambda-\lambda\pri)T}}{i(\lambda-\lambda\pri)T}}\\
\leq&\sum_{\lambda\neq\lambda\pri}\size{\braket{\lambda}{\text{\sc entrance}}}^2\size{\braket{\lambda\pri}{\text{\sc entrance}}}^2\frac{1-e^{-i(\lambda-\lambda\pri)T}}{i(\lambda-\lambda\pri)T}\\
\leq&\frac{2}{T\delta}
\end{flalign}
where $\delta$ is the difference between the smallest gap between any two distinct eigenvalues of $A$.  As a result, we get
\be\Pr(\text{\sc entrance},\text{\sc exit}, T)\geq\frac{1}{2d-1}-\frac{2}{T\delta}.\ts\ee
Childs et al. \cite{780552} show that $\delta$ is $\Omega(1/d^3)$.  Therefore, if we take $T$ of size $O(d^4)$, we get success probability $O(1/d)$.  Repeating this process, we can achieve an arbitrarily high probability of success in time polynomial in $d$--- an exponential speedup over the classical random walk.

\subsubsection{$\AND$-$\OR$ Tree Evaluation}
$\AND$-$\OR$ trees arise naturally when evaluating the value of a two player combinatorial game.  We will call the players $P_0$ and $P_1$.  The positions in this game are represented by nodes on a tree.  The game begins at the root, and players alternate, moving from the root towards the leaves of the tree.  For simplicity, we assume that we are dealing with a binary tree; that is, for each move, the players have exactly two moves.  While the algorithm can be generalizes to any approximately balanced tree, but we consider the binary case for simplicity.  We also assume that every game lasts some fixed number of turns $d$, where a turn consists of one player making a move.  We denote the total number of leaf nodes by $n=2^d$.  We can label the leaf nodes according to which player wins if the game reaches each node; they are labeled with a $0$ if $P_0$ wins, and a $1$ if $P_1$ wins.  We can then label each node in the graph by considering its children.  If a node corresponds to $P_0$'s turn, we take the $\AND$ of the children.  For $P_1$'s move, we take the $\OR$ of the children.  The value at the root node tells us which player has a winning strategy, assuming perfect play.

\begin{figure}[htp]
\sidecaption
\includegraphics[scale=.7]{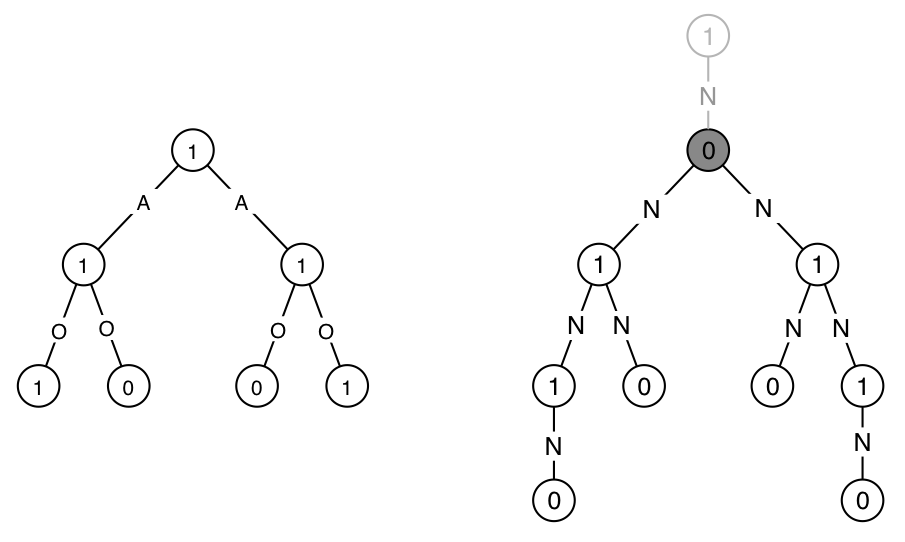}
\caption{An $\AND$-$\OR$ tree and the equivalent $\NAND$ tree, with example values.  Note that the top-most node will be omitted from now on, and the shaded node in the $\NAND$ tree will be referred to as the {\sc root} node.}
\label{fig:nandtree}
\end{figure}

Now, since
\begin{flalign}
\AND(x_1,...,x_k)=&\NAND(\NAND(x_1,...,x_k))\\
\OR(x_1,...,x_k)=&\NAND(\NAND(x_1),...,\NAND(x_k))
\end{flalign}
we can rewrite the $\AND$-$\OR$ tree using only $\NAND$ operations.  Furthermore, since
\be
\NOT(x_1)=\NAND(x_1)
\ee
rather than label leaves with a $1$, we can insert an extra node below the leaf and label it $0$.  Also, consecutive $\NAND$ operations cancel out, and will be removed.  This is illustrated in figure \ref{fig:nandtree}.  Note that the top-most node will be omitted from now on, and the shaded node will be referred to as the {\sc root} node.  Now, the entire $\AND$-$\OR$ tree is encoded in the structure of the $\NAND$ tree.  We will write $\NAND(v)$ to denote the $\NAND$ of the value obtained by evaluating the tree up to a vertex $v$.  The value of the tree is therefore $\NAND(\text{{\sc root}})$.

The idea of the algorithm is to use the adjacency matrix of the $\NAND$ tree as the Hamiltonian for a continuous time quantum walk.  We will show that the eigenspectrum of this walk operator is related to the value of $\NAND(\text{{\sc root}})$.  This idea first appeared in a paper of Farhi, Goldman and Gutmann \cite{arXiv:quant-ph/0702144}, and was later refined by Ambainis, Childs and Reichardt \cite{4389507}.  We begin with the following lemma:
\begin{lemma} Let $\ket{\lambda}$ be an eigenvector of $A$ with value $\lambda$.  Then, if $v$ is a node in the $\NAND$ tree with parent node $p$ and children $C$,
\be \braket{p}{\lambda}=-\lambda\braket{v}{\lambda}+\sum_{c\in C}\braket{c}{\lambda}\ee
\end{lemma}
Therefore, if $A$ has an eigenvector with value $0$, then \be \braket{p}{\lambda_0}=\sum_{c\in C}\braket{c}{\lambda_0}\ee
Using this fact and some inductive arguments, Ambainis, Childs and Reichardt  \cite{4389507} prove the following theorem:
\begin{theorem} If $\NAND(\text{{\sc root}})=0$, then there exists an eigenvector $\ket{\lambda_0}$ of $A$ with eigenvalue $0$ such that $\size{\braket{\text{{\sc root}}}{\lambda_0}}\geq\oort$.  Otherwise, if $\NAND(\text{{\sc root}})=1$, then for any eigenvector $\ket{\lambda}$ of $A$ with $\bra{\lambda}A\ket{\lambda}<\frac{1}{2\sqrt{n}}$, we have $\braket{\text{{\sc root}}}{\lambda_0}=0$.
\end{theorem}
This result immediately leads to an algorithm.  We perform phase estimation with precision $O(1/\sqrt{n})$ on the quantum walk, beginning in the state $\ket{\text{{\sc root}}}$.  If $\NAND(\text{{\sc root}})=0$, get a phase of $0$ with probability $\geq\half$.  If $\NAND(\text{{\sc root}})=1$, we will never get a phase of $0$.  This gives us a running time of $O(\sqrt{n})$.  While we have restricted our attention to binary trees here, this result can be generalized to any $m$-ary tree.

It is worth noting that unstructured search is equivalent to taking the $\OR$ of $n$ variables.  This is a $\AND$-$\OR$ tree of depth 1, and $n$ leaves.  The $O(\sqrt{n})$ running time of Grover's algorithm corresponds to the running time for the quantum walk algorithm.

Classically, the running time depends not just on the number of leaves, but on the structure of the tree.  If it is a balanced $m$-ary tree of depth $d$, then the running time is \be O\left(\left(\frac{m-1+\sqrt{m^2+14m+1}}{4}\right)^d\right).\ts\ee In fact, the quantum speedup is maximal when we have an $n$-ary tree of depth $1$.  This is just unstructured search, and requires $\Omega(n)$ time classically.

Reichardt and \u{S}palek \cite{1374394} generalize the AND-OR tree problem to the evaluation of a broader class of logical formulas.  Their approach uses {\em span programs}.  A span program $P$ consists of a set of target vector $t$, and a set of input vectors $\{v_1^0,v_1^1,...,v_n^0,v_n^1,\}$ corresponding to logical literals $\{x_1,\overline{x_1},...,x_n,\overline{x_n}\}$.  The program corresponds to a boolean function $f_P:\{0,1\}^n\to\{0,1\}$ such that, for $\sigma\in\{0,1\}^n$, $f(\sigma)=1$ if and only if $$\sum_{j=1}^nv_j^{\sigma_j}=t.\qquad\text{\cite{Karchmer:1993sf}}$$Reichardt and \u{S}palek outline the connection between span programs and the evaluation of logical formulas.  They show that finding $\sigma\in f^{-1}(1)$ is equivalent to finding a zero eigenvector for a graph $G_P$ corresponding to the span program $P$.  In this sense, the span program approach is similar to the quantum walk approach of Childs et al.--- both methods evaluate a formula by finding a zero eigenvector of a corresponding graph.

\section{Tensor Networks and Their Applications}\label{sec:tensor-networks}

A tensor network consists of an underlying graph $G$, with an algebraic object called a {\em tensor} assigned to each vertex of $G$.  The value of the tensor network is calculated by performing a series of operations on the associated tensors.  The nature of these operations is dictated by the structure of $G$.  At their simplest, tensor networks capture basic algebraic operations, such as matrix multiplication and the scalar product of vectors.  However, their underlying graph structure makes them powerful tools for describing combinatorial problems as well.  We will explore two such examples--- the Tutte polynomial of a planar graph, and the partition function of a statistical mechanical model defined on a graph.  As a result, the approximation algorithm that we will describe below for the value of a tensor network is implicitly an algorithm for approximating the Tutte polynomial, as well as the partition function for these statistical mechanics models.  We begin by defining the notion of a tensor.  We then outline how these tensors and tensor operations are associated with an underlying graph structure.  For a more detailed account of this algorithm, the reader is referred to \cite{arXiv:0805.0040}.  Finally, we will describe the quantum approxiamtion algorithm for the value of a tensor network, as well as the applications mentioned above.

\subsubsection{Tensors: Basic Definitions}
Tensors are formally defined as follows:
\begin{definition} A tensor $M$ of rank $m$ and dimension $q$ is an element of $\mathbf{C}^{q^m}$.  Its entries are denoted by $M_{j_1,j_2,...,j_m}$, where $0\leq j_k\leq q-1$ for all $j_k$.\end{definition}
Based on this definition, a vector is simply a tensor of rank 1, while a square matrix is a tensor of rank 2.  We will now define several operations on tensors, which will generalize many familiar operations from linear algebra.
\begin{definition} Let $M$ and $N$ be two tensors of dimension $q$ and rank $m$ and $n$ respectively.  Then, their product, denoted $M\otimes N$, is a rank $m+n$ tensor with entries \be(M\otimes N)_{j_1,...,j_m,k_1,...,k_n}=M_{j_1,...,j_m}\cdot N_{k_1,...,k_n}.\ee\end{definition}
This operation is simply the familiar tensor product.   While the way that the entries are indexed is different, the resulting entries are the same.
\begin{definition} Let $M$ be a tensor of rank $m$ and dimension $q$.  Now, take $a$ and $b$ with $1\leq a <b\leq m$.  The contraction of $M$ with respect to $a$ and $b$ is a rank $m-2$ tensor $N$ defined as follows:
\be N_{j_1,...,j_{a-1},j_{a+1},...j_{b-1},j_{b+1},...,j_m}=\sum_{k=0}^{q-1}M_{j_1,...,j_{a-1},k,j_{a+1},...j_{b-1},k,j_{b+1},...,j_m}.\ee
\end{definition}
One way of describing this operation is that each entry in the contracted tensor is given by summing along the ``diagonal" defined by $a$ and $b$.  This operation generalizes the partial trace of a density operator.  The density operator of two qubit system can be thought of as a rank 4 tensor of dimension 2.  Tracing out the second qubit is then just taking a contraction with respect to 3 and 4.  It is also useful to consider the combination of these two operations:
\begin{definition} If $M$ and $N$ are two tensors of dimension $q$ and rank $m$ and $n$, then for $a$ and $b$ with $1\leq a\leq m$ and $1\leq b\leq n$, the contraction of $M$ and $N$ is the result of contracting the product $M\otimes N$ with respect to $a$ and $m+b$.\end{definition}
We now have the tools to describe a number of familiar operations in terms of tensor operations.  For example, the inner product of two vectors can be expressed as the contraction of two rank 1 tensors.  Matrix multiplication is just the contraction of 2 rank 2 tensors $M$ and $N$ with respect to the second index of $M$ and the first index of $N$.  Finally, if we take a Hilbert space $H=\mathbf{C}^q$, then we can identify a tensor $M$ of dimension $q$ and rank $m$ with a linear operator $M^{s,t}:H^{\otimes t}\rightarrow H^{\otimes s}$ where $s+t=m$: \be \label{Mst}M^{s,t}=\sum_{j_1,...,j_m}M_{j_1,...,j_m}\ket{j_1}\otimes...\otimes\ket{j_s}\bra{j_{s+1}}\otimes...\otimes\bra{j_m}.\ee
This correspondence with linear operators is essential to understanding tensor networks and their evaluation.

\subsection{The Tensor Network Representation}
\begin{figure}[htp]
\sidecaption
\includegraphics[scale=.7]{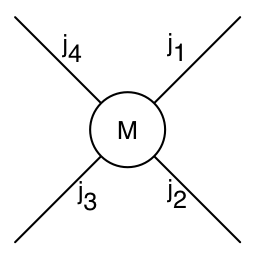}
\qquad\qquad\includegraphics[scale=.7]{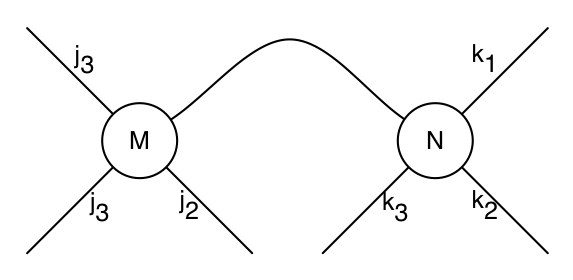}
\caption{The network representation of a tensor $M$ of rank 4, and the contraction of two rank 4 tensors, $M$ and $N$.}
\label{fig:tensorops}
\end{figure}

A tensor network $T(G,\mathbf{M})$ consists of a graph $G=(V,E)$ and a set of tensors $\mathbf{M}=\{M[v]: v\in V\}$.  A tensor is assigned to each vertex, and the structure of the graph $G$ encodes the operations to be performed on the tensors.  We say that the {\em value} of a tensor network is the tensor that results from applying the operations encoded by $G$ to the set of tensors $\mathbf{M}$.  When the context is clear, we will let $T(G,\mathbf{M})$ denote the value of the network.  In addition to the typical features of a graph, we allow $G$ to have {\em free edges}--- edges that are incident with a vertex on one end, but are unattached at the other.  For simplicity, we will assume that all of the tensors in $\mathbf{M}$ have the same dimension $q$.  We also require that $deg(v)$, the degree of vertex $v$, is equal to the rank of the associated tensor $M[v]$.  If $G$ consists of a single vertex $v$ with $m$ free edges, then $T(G,\mathbf{M})$ represents a single tensor of rank $m$.  Each index of $M$ is associated with a particular edge.  It will often be convenient to refer to the edge associated with an index $a$ as $e_a$.  We now define a set of rules that will allow us to construct a tensor network corresponding to any sequence of tensor operations:
\begin{itemize}
\item If $M$ and $N$ are the value of two tensor networks $T(G,\mathbf{M})$ and $T(H,\mathbf{N})$, then taking the product $M\otimes N$ is equivalent to taking the disjoint union of the two tensor networks, $T(G\cup H, \mathbf{M}\cup\mathbf{N})$
\item If $M$ is the value of a tensor network $T(G,\mathbf{M})$, then contracting $M$ with respect to $a$ and $b$ is equivalent to joining the free edges $e_a$ and $e_b$ in $G$.
\item As a result, taking the contraction of $M$ and $N$ with respect to $a$ and $b$ corresponds to joining the associated edge $e_a$ from $T(G,\mathbf{M})$ and $e_b$ from $T(H,\mathbf{N})$
\end{itemize}
Some examples are illustrated in figure \ref{fig:tensorops}.  Applying these simple operations iteratively allows us to construct a network corresponding to any series of products and contractions of tensors.  Note that the number of free edges in $T(G,\mathbf{M})$ is always equal to the rank of the corresponding tensor $M$.  We will be particularly interested in tensor networks that have no free edges, and whose value is therefore a scalar.

\begin{figure}[htp]
\centering
\includegraphics[scale=.7]{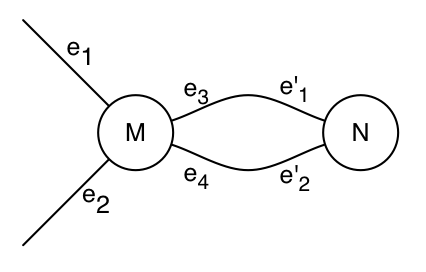}
\caption{The network corresponding to the operator $M^{2,2}$ acting on a rank 2 tensor $N$.}
\label{fig:operatornetwork}
\end{figure}

We can now consider the tensor network interpretation of a linear operator $M^{s,t}$ defined in (\ref{Mst}).  $M^{s,t}$ is associated with a rank $m=s+t$ tensor $M$.  An equivalent tensor network could consist of a single vertex with $m$ free edges $(e_1,...,e_m)$.  The operator $M^{s,t}$ acts on elements of $(\mathbf{C}^q)^{\otimes t}$, which are rank $t$ tensors, and can therefore be represented a single vertex with $t$ free edges $(e\pri_1,...,e\pri_t)$.  The action of $M^{s,t}$ on  an element $N\in(\mathbf{C}^q)^{\otimes t}$ is therefore represented by connecting $e\pri_k$ to $e_{s+k}$ for $k=1,...,t$.  Figure \ref{fig:operatornetwork} illustrates the operator $M^{2,2}$ acting on a rank 2 tensor.  Note that the resulting network has two free edges, corresponding to the rank of the resulting tensor.

We now return to the tensor networks we will be most interested in--- those with no free edges.  Let us first consider the example of an inner product of two rank 1 tensors.  This corresponds to a graph consisting of two vertices $u$ and $v$ joined by a single edge.  The value of the tensor is then \be T(G,\mathbf{M})=\sum_{j=0}^q(M_u)_j\cdot(M[v])_j\ee  That is, it is a sum of $q$ terms, each of which corresponds to the assignment of an element of $\{0,...,q-1\}$ to the edge $(u,v)$.  We can refer to this assignment as an $q$-edge colouring of the graph.  In the same way, the value of a more complex tensor network is given by a sum whose terms correspond to $q$-edge colourings of $G$.  Given an colouring of $G$, let $M[v]^\gamma=M[v]_{i_1,...,i_m}$ where $i_1,...,i_m$ are the values assigned by $\gamma$ to the edges incident at $v$.  That is, a $q$-edge colouring specifies a particular entry of each tensor in the network.  Then, we can rewrite the value of $T(G,\mathbf{M})$ as follows:
\be T(G,\mathbf{M})=\sum_\gamma\left(\prod_{v\in V}M[v]^\gamma\right)\ee
where $\gamma$ runs over all $q$-edge colourings of $G$.  Thinking of the value of a tensor network as a sum over all $q$-edge colourings of $G$ will prove useful when we consider applications of tensor networks to statistical mechanics models.
\subsection{Evaluating Tensor Networks: the Algorithm}
In this section, we will first show how to interpret a tensor network as a series of linear operators.  We will then show how to apply these operators using a quantum circuit.  Finally, we will see how to approximate the value of the tensor network using the Hadamard test.
\subsubsection{Overview}
\begin{figure}[htp]
\sidecaption
\includegraphics[scale=1]{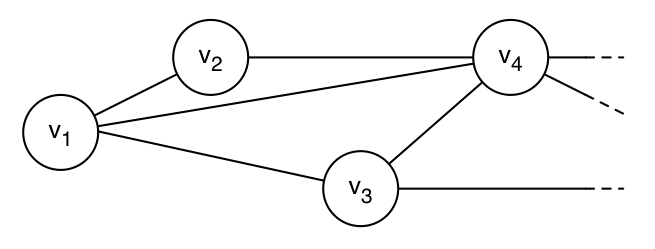}
\caption{Part of the graph $G$, with an ordering on $V(G)$.}
\label{fig:vertexordering}
\end{figure}

In order to evaluate a tensor network $T(G,\mathbf{M})$, we first give it some structure, by imposing an ordering on the vertices of $G$, $$V(G)=\{v_1,...,v_n\}.$$We further define the sets $S_j=\{v_1,...,v_j\}$.  This gives us a natural way to depict the tensor network, which is shown in figure \ref{fig:vertexordering}.  Our evaluation algorithm will ``move" from left to right, applying linear operators corresponding to the tensors $M_{v_i}$.  We can think of the edges in analogy to the ``wires" of a quantum circuit diagram.
To make this more precise, let $v_i$ be a vertex of degree $d$.  Let $t$ be the number of vertices connecting $v_i$ to $S_{i-1}$ and $s$ be the number of edges connecting $v_i$ to $V(G)\setminus S_i$.  Then, the operator that is applied at $v_i$ is $M_{v_i}^{s,t}$.  Letting $j$ be the number of edges from $S_{i-1}$ to $V(G)\setminus S_i$, we apply the identity operator $\mathbf{I}_j$ on the $j$ indices that correspond to edges not incident at $v_i$.  Combining these, we get an operator \be \label{Ni} N_i=M_{v_i}^{s,t}\otimes\mathbf{I}_j.\ee Note that, taking the product of these operators, $\prod_{i=1}^nN_i$, gives the value of $T(G,\mathbf{M})$.  However, there are a few significant details remaining.  Most notably, the operators $N_i$ are not, in general, unitary.  In fact, they may not even correspond to square matrices.  We will now outline a method for converting the $N_i$ into equivalent unitary operators.

\subsubsection{Creating Unitary Operators}
Our first task is to convert the operators $N_i$ into ``square" operators--- that is, operators whose domain and range have the same dimension.  Referring to (\ref{Ni}), the identity $I_j$ is already square, so we need only modify $M_{v_i}^{s,t}$.  In order to do this, we will add some extra qubits in the $\ket0$ state.  We consider three cases:
\begin{enumerate}
\item If $s=t$, we define $\widetilde{M_{v_i}^{s,t}}=M_{v_i}^{s,t}$
\item If $s>t$, then we add $s-t$ extra qubits in the $\ket0$ state, and define $\widetilde{M_{v_i}^{s,t}}$ such that \be\widetilde{M_{v_i}^{s,t}}\bigl(\ket{\psi}\otimes\ket0^{\otimes(s-t)}\bigr)=M_{v_i}^{s,t}\ket\psi.\ee
For completeness, we say that if the last $s-t$ qubits are not in the $\ket0$ state, $\widetilde{M_{v_i}^{s,t}}$ takes them to the 0 vector.
\item If $s<t$, then we define \be\widetilde{M_{v_i}^{s,t}}\ket{\psi}=(M_{v_i}^{s,t}\ket\psi)\otimes\ket0^{\otimes(t-s)}.\ee
\end{enumerate}
Finally, we define \be \tilde{N}_i=\widetilde{M_{v_i}^{s,t}}\otimes\mathbf{I}_j,\ee which is a square operator.  Now, to derive corresponding unitary operators, we make use of the following lemma:
\begin{lemma} \label{unitarylemma}Let $M$ be a linear map $A:H^{\otimes t}\to H^{\otimes t}$, where $H$ is a Hibert space $H=\mathbf{C}^q$.  Furthermore, let $G=\mathbf{C}^2$ be a spanned by $\{\ket0,\ket1\}$.  Then, there exists a unitary operator $U_M:\left(H^{\otimes t}\otimes G\right)\to \left(H^{\otimes t}\otimes G\right)$ such that \be U_M(\ket{\psi}\otimes\ket0)=\frac{1}{\vnorm{M}}M\ket\psi\otimes\ket0+\ket\phi\otimes\ket1\ee $U_M$ can be implemented on a quantum computer in $\text{poly}(q^t)$ time.\end{lemma}
A proof can be found in \cite{arXiv:0805.0040} as well as \cite{arXiv:quant-ph/0702008}.
Applying this lemma, we can create $n$ unitary operators $U_{\tilde{N}_j}$ for $1\leq j\leq n$, and define \be U=\prod_{j=1}^nU_{\tilde{N}_j}.\ee
It is easily verified that
\be\label{innerp}\bra{0}^{\otimes r}U\ket{0}^{\otimes r}=\frac{T(G,\mathbf{M})}{\prod_j\vnorm{M_{v_j}^{s,t}}}\ee where $r$ is dependent on the number of vertices $n$ as well as the structure of $G$ and the ordering of the vertices $v_1,...,v_n$.  

\subsubsection{The Hadamard Test} \label{sec:hadamard-test}
In order to approximate $\bra{0}^{\otimes r}U\ket{0}^{\otimes r}$, most authors suggest the {\em Hadamard test}--- a well-known method for approximating the weighted trace of a unitary operator.  First, we add an ancillary qubit that will act as a control register.  We then apply the circuit outlined below, and measure the ancillary qubit in the computational basis.

\begin{figure}[htp]
{\Large{$$\Qcircuit @C=1.5em @R=1.5em @!C {&\lstick{\ket{0}}&\gate{H} &\ctrl{1}&\gate{H}&\qw \\&\lstick{\ket{\psi}}& \qw&\gate{U}&\qw&\qw}$$}}
\caption{The quantum circuit for the Hadamard test.}
\label{fig:hadamard-test}
\end{figure}
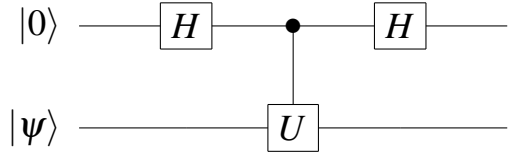

It is not difficult to show that we measure
\begin{flalign*}\ket{0}&\text{ with probability }\frac{1}{2}\bigl(1+\text{Re}\bra\phi U\ket\psi\bigr)\\
\ket{1}&\text{ with probability }\frac{1}{2}\bigl(1-\text{Re}\bra\psi U\ket\psi\bigr).
\end{flalign*}
So, if we assign a random variable $X$ such that $X=1$ when we measure $\ket0$ and $X=-1$ when we measure $\ket1$, then $X$ has expected value $\text{Re}\bra\psi U\ket\psi$.  So, in order to approximate $\text{Re}\bra\psi U\ket\psi$ to a precision of $\epsilon$ with constant probability $p>1/2$, we require $O(\epsilon^{-2})$ repetitions of this protocol.  In order to calculate the imaginary portion, we apply the gate $$R=\tbt{1}{0}{0}{-i}$$ to the ancillary qubit, right after the first Hadamard gate.  It is important to note that the Hadamard test gives an {\em additive approximation} of $\bra{0}^{\otimes r}U\ket{0}^{\otimes r}$.  Additive approximations will be examined in section \ref{sec:addapprox}.

\subsubsection{Approximating $\bra{0}^{\otimes r}U\ket{0}^{\otimes r}$ Using Amplitude Estimation}
In order to estimate $\bra{0}^{\otimes r}U\ket{0}^{\otimes r}$, most of the literature applies the Hadamard test.  However, it is worth noting that we can improve the running time by using the process of {\em amplitude estimation}, as outlined in section \ref{sec:amplitude-amplification}.  Using the notation from section \ref{sec:amplitude-amplification}, we have $U_f=U_0$ and $A=U$, the unitary corresponding to the tensor network $T(G,\mathbf{M})$.  We begin in the state $U\ket0^{\otimes r}$, and our search iterate is \be Q=-UU_0U^{-1}U_0.\ee In order to approximate $\bra{0}^{\otimes r}U\ket{0}^{\otimes r}$ to a precision of $\epsilon$, our running time is in $O(1/\epsilon)$, a quadratic improvement over the Hadamard test.

\subsubsection{Additive Approximation}\label{sec:addapprox}
Let $f:S\to\mathbf{C}$ be a function that we would like to evaluate.  Then, an {\em additive approximation} $A$ with approximation scale $\Delta:S\to\mathbf{R}$ is an algorithm that, on input $x\in S$ and parameter $\epsilon>0$ , outputs $A(x)$ such that
\be \text{Pr}\Bigl(\size{A(x)-f(x)}\geq\epsilon\Delta(x)\Bigr)\leq c\ee
for some constant $c$ with $0\leq c<1/2$.  If $\Delta(x)$ is $O(f(x))$, then $A$ is a {\em fully polynomial randomized approximation scheme}, or FPRAS.

We will now determine the approximation scale for the algorithm outlined above.  Using amplitude estimation, we estimate
\be\bra{0}^{\otimes r}U\ket{0}^{\otimes r}=\frac{T(G,\mathbf{M})}{\prod_j\vnorm{M_{v_j}^{s,t}}}\ee
to within $\epsilon$ with time requirement in $O(1/\epsilon)$.  However, the quantity we actually want to evaluate is $T(G,\mathbf{M})$, so we must multiply by $\prod_j\vnorm{M_{v_j}^{s,t}}$.  Therefore, our algorithm has approximation scale
\be \Delta(G,\mathbf{M})=\prod_j\vnorm{M_{v_j}^{s,t}}\ee
We apply lemma \ref{unitarylemma} to each vertex in $G$, and require $O(1/\epsilon)$ repetitions of the algorithm to approximate $T(G,\mathbf{M})$ with the desired accuracy using amplitude estimation.  This gives an overall running time \be O\left(\text{poly}(q^{D(G)})\cdot\size{V(G)}\cdot\epsilon\right)\ee where $D(G)$ is the maximum degree of any vertex in $G$.  The algorithm is, therefore, an additive approximation of $T(G,\mathbf{M})$.
\subsection{Approximating the Tutte Polynomial for Planar Graphs Using Tensor Networks}
In 2000, Kitaev, Freedman, Larsen and Wang \cite{arXiv:quant-ph/0001071,arXiv:quant-ph/0101025} demonstrated an efficient quantum simulation for topological quantum field theories.  In doing so, they implied that the Jones polynomial can be efficiently approximated at $e^{2\pi i/5}$.  This suggests that efficient quantum approximation algorithms might exist for a wider range of knot invariants and values.  Aharonov, Jones and Landau developed such a Jones polynomial for any complex value \cite{arXiv:quant-ph/0511096}.  Yard and Wocjan's developed a related algorithm for approximating the HOMFLYPT polynomial of a braid closure \cite{arXiv:quant-ph/0603069}.  Both of these knot invariants are special cases of the Tutte polynomial.  While the tensor network algorithm in section \ref{sec:tuttepoly} follows directly from a later paper of Aharonov et al. \cite{arXiv:quant-ph/0702008}, it owes a good deal to these earlier results for knot invariants.

We will give a definition of the Tutte polynomial, as well as an overview of its relationship to tensor networks and the resulting approximation algorithm.

\subsubsection{The Tutte Polynomial}
The multi-variate Tutte Polynomial is defined for a graph $G=(V,E)$ with edge weights $\mathbf{w}=\{w_e\}$ and variable $q$ as follows:
\be Z_G(q,\mathbf{w})=\sum_{A\subseteq E}q^{k(A)}\prod_{e\in A}w_e\ee
where $k(A)$ denotes the number of connected components in the graph induced by $A$.  The power of the Tutte polynomial arises from the fact that it captures nearly all functions on graphs defined by a {\em skein relation}.  A skein relation is of the form
\be f(G)=x\cdot f(G/e)+y\cdot f(G\setminus e)\ee
where $G/e$ is created by contracting an edge $e$ and $G\setminus e$ is created by deleting $e$.  Oxley and Welsh \cite{9231116} show that, with a few additional restrictions on $f$, if $f$ is defined by a skein relation, then computing $f$ can be reduced to computing $Z_G$.  It turns out that many functions can be defined in terms of a skein relation, including
\begin{enumerate}
\item The partition functions of the Ising and Potts models from statistical physics.
\item The chromatic and flow polynomials of a graph $G$.
\item The Jones Polynomial of an alternating link.
\end{enumerate}
The exact evaluation of the Tutte polynomial, even when restricted to planar graphs, turns out to be $\#P$-hard for all but a handful of values of $q$ and $\mathbf{w}$.  An exact algorithm is therefore believed to be impossible for a quantum computer.  Indeed, even a polynomial time FPRAS  seems very unlikely for the Tutte polynomial of a general graph and any parameters $q$ and ${\bf w}$.  The interesting question seems to be {\em which} graphs and parameters admit an efficient and accurate approximation.  By describing the Tutte polynomial as the evaluation of a Tensor network, we immediately produce an additive approximation algorithm for the Tutte polynomial.  However, we do not have a complete characterization of the graphs and parameters for which this approximation is non-trivial.
\subsubsection{The Tutte Polynomial as a Tensor Network}\label{sec:tuttepoly}
\begin{figure}[htp]
\includegraphics[scale=1.5]{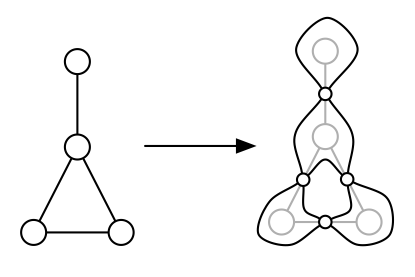}
\caption{A graph $G$ and the resulting medial graph $L_G$.}
\label{fig:medial}
\end{figure}

Given a planar graph $G$, and an embedding of $G$ in the plane, we define the {\em medial graph} $L_G$.  The vertices of $L_G$ are placed on the edges of $G$.  Two vertices of $L_G$ are joined by an edge if they are adjacent on the boundary of a face of $G$.  If $G$ contains vertices of degree 2, then $L_G$ will have multiple edges between some pair of vertices.  Also note that $L_G$ is a regular graph with valency 4.  An example of a graph and its associated medial graph is depicted in figure \ref{fig:medial}.

\begin{figure}[htp]
\sidecaption
\includegraphics[scale=.84]{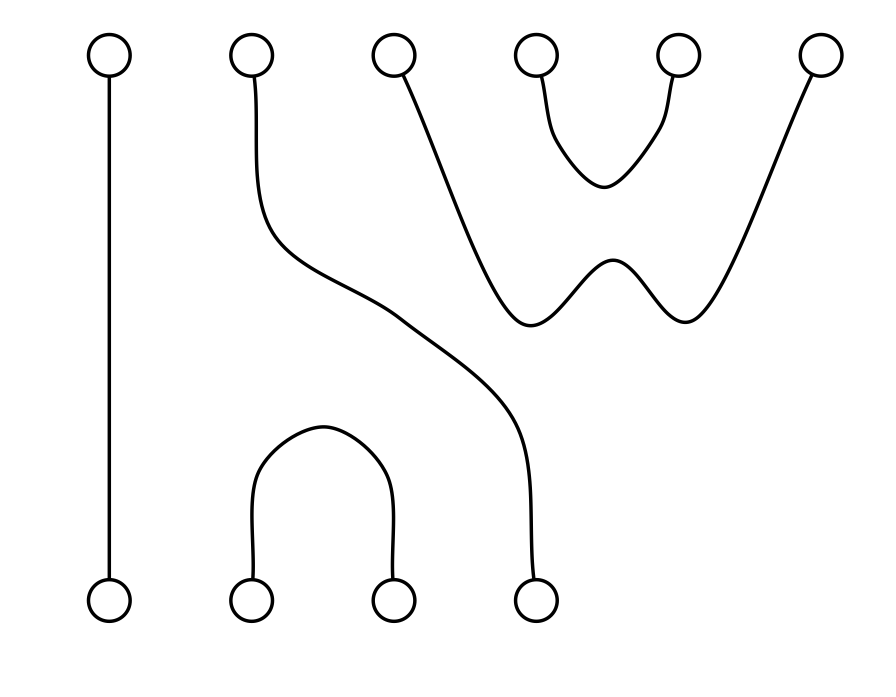}
\caption{An element of the Temperley-Lieb algebra with 6 upper pegs and 4 lower pegs.}
\label{fig:GTL}
\end{figure}

In order to describe the Tutte polynomial in terms of a tensor network, we make use of the {\em generalized Temperley-Lieb algebra}, as defined in \cite{arXiv:quant-ph/0702008}.  The basis elements of the Temperley-Lieb algebra $GTL(d)$ can be thought of as diagrams in which $m$ upper pegs are joined to $n$ lower pegs by a series of strands.  The diagram must contain no crossings or loops.  See figure \ref{fig:GTL} for an example of such an element.  Two basis elements are considered equivalent if they are isotopic to each other or if one can be obtained from the other by `padding' it on the right with a series of vertical strands.  The algebra consists of all complex weighted sums of these basis elements.  Given two elements of the algebra $T_1$ and $T_2$, we can take their product $T_2\cdot T_1$ by simply placing $T_2$ on top of $T_1$ and joining the strands.  Note that if the number of strands do not match, we can simply pad the appropriate element with some number of vertical strands.  As a consequence, the identity element consists of any number of vertical strands.
In composing two elements in this way, we may create a loop.  Let us say that $T_1\cdot T_2$ contains one loop.  Then, if $T_3$ is the element created by removing the loop, we define $T_1\cdot T_2=dT_3$, where $d$ is the complex parameter of $GTL(d)$.

\begin{figure}[htp]
\sidecaption
\includegraphics[scale=.65]{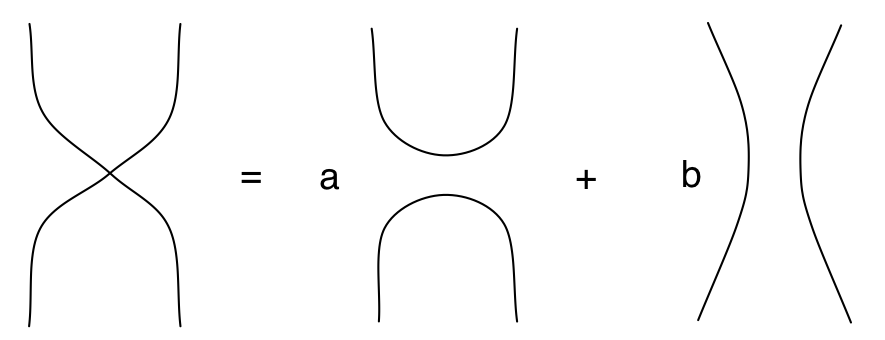}
\caption{An element that contains a crossing is equated to a weighted sum of the elements obtained by `opening' the crossing.}
\label{fig:opening1}
\end{figure}

We would also like this algebra to accommodate drawings with crossings.  Let $T_1$ be such a diagram, and $T_2$ and $T_3$ be the diagrams resulting from `opening' the crossing in the two ways indicated in figure \ref{fig:opening1}.  Then, we define $T_1=aT_2+bT_3$, for appropriately defined $a$ and $b$.

If we think of $L_G$ as the projection of a knot onto the plane, where the vertices of $L_G$ are the crossings of the knot, we see that $L_G$ can be expressed in terms of the generalized Temperly-Lieb algebra.  We would like to take advantage of this in order to map $L_G$ to a series of tensors that will allow us to approximate the Tutte polynomial of $G$.  Aharonov et. al. define such a representation $\rho$ of the Temperley Lieb algebra.  If $T\in GTL(d)$ is a basis element with $m$ lower pegs and $n$ upper pegs, then $\rho$ is a linear operator such that  \be\rho(T):H^{\otimes m+1}\to H^{\otimes n+1}\ee where $H=\mathbf{C}^k$ for some $k$ (which depends on the particular embedding of $G$) such that
\begin{enumerate}
\item $\rho$ preserves multiplicative structure.  That is, if $T_1$ has $m$ upper pegs and $T_2$ has $m$ lower pegs, then \be\rho(T_2\cdot T_1)=\rho(T_2)\cdot\rho(T_1).\ee
\item $\rho$ is linear.  That is, \be\rho(\alpha T_1+\beta T_2)=\alpha \rho(T_1)+\beta \rho(T_2).\ee
\end{enumerate}
For our purposes, $k$ can be upper bounded by the number of edges $\size{E(L_G)}=2\size{E(G)}$, and corresponds to the value $r$ in equation (\ref{innerp}).  To represent $L_G$, each crossing (vertex) of $L_G$ is associated with a weighted sum of two basis elements, and therefore is represented by a weighted sum of the corresponding linear operators, $\rho(T_1)=a\rho(T_2)+b\rho(T_3)$.  The minima and maxima of $L_G$ are also basis elements of $GTL(d)$, and can be represented as well.  Finally, since $L_G$ is a closed loop, and has no `loose ends', $\rho(L_G)$ must be a scalar multiple of the identity operator on $H=\mathbf{C}^k$.  This relates well to our notion that the value of a tensor network with no loose ends is just a scalar.

We would like this scalar to be the Tutte polynomial of the graph $G$.  In order to do this, we need to choose the correct values for $a$, $b$ and $d$.  Aharonov et. al. show how to choose these values so that the scalar is a graph invariant called the Kauffman bracket, from which we can calculate the Tutte polynomial for planar graphs.  So, we can construct a tensor network whose value is the Kauffman bracket of $L_G$, and thus gives us the Tutte polynomial of $G$ by the following procudure:
\begin{enumerate}
\item Construct the medial graph $L_G$ and embed it in the plane.
\item Let $L_G\pri$ be constructed by adding a vertex at each local minimum and maximum of $L_G$.  We need these vertices because we need to assign the linear operator associated with each minimum/maximum with a vertex in the tensor network.
\item Each vertex $v$ of $L_G\pri$ has a Temperley-Lieb element $T_v$ associated with it.  For some vertices, this is a crossing; for others, it is a local minimum or maximum.  Assign the tensor $M[v]=\rho(T_v)$ to the vertex $v$.
\end{enumerate}
The tensor network we are interested in is then $T(L_G\pri,\mathbf{M})$, where $\mathbf{M}=\{M[v]: v\in V(L_G\pri)\}$.  Approximating the value of this tensor network gives us an approximation of the Kauffman bracket of $L_G$, and therefore of the Tutte polynomial of $G$.  Furthermore, we know that $L_G\pri$ has maximum degree 4.  Finally, we will assume that  giving a running time of
 \be O\left(\text{poly}(q^4)\cdot\size{E(G)}\cdot\epsilon^{-1}\right)\ee
 In this case, $q$ depends on the particular embedding f $L_G$, but can be upper bounded by $\size{E}$.  For more details on the representation $\rho$ and quantum approximations of the Tutte polynomial as well as the related Jones polynomial, the reader is referred to the work of Aharonov et al., in particular \cite{arXiv:quant-ph/0605181} and \cite{arXiv:quant-ph/0702008}.

\subsection{Tensor Networks and Statistical Mechanics Models}
Many models from statistical physics describe how simple local interactions between microscopic particles give rise to macroscopic behaviour.  See \cite{arXiv:0804.2468} for an excellent introduction to the combinatorial aspects of these models.  The models we are concerned with here are described by a weighted graph $G=(V,E)$.  In this graph, the vertices represent particles, while edges represent an interaction between particles.  A configuration $\sigma$ of a $q$-state model is an assignment of a value from the set $\{0,...,q-1\}$ to each vertex of $G$.  We denote the value assigned to $v$ by $\sigma_v$.  For each edge $e=(u,v)$, we define a local Hamiltonian $h_e(\sigma_u,\sigma_v)\in\mathbb{C}$.  The Hamiltonian for the entire system is then given taking the sum:
\be H(\sigma)=\sum_{e=(u,v)}h_e(\sigma_u, \sigma_v)\ee The sum runs over all the edges of $G$.  The {\em partition function} is then defined as
\be Z_G(\beta)=\sum_\sigma e^{-\beta H(\sigma)}\ee
where $\beta=1/kT$ is referred to as the {\em inverse temperature} and $k$ is Boltzmann's constant.  The partition function is critical to understanding the behaviour of the system.  Firstly, the partition function allows us to determine the probability of finding the system in a particular configuration, given an inverse temperature $\beta$:
\be\Pr(\sigma\pri)=\frac{e^{-\beta H(\sigma\pri)}}{Z_G(\beta)}\ee This probability distribution is known as the Boltmann distribution.  Calculating the partition function also allows us to derive other properties of the system, such as entropy and free energy.  An excellent discussion of the partition function from a combinatorial perspective can be found in \cite{9231116}.

\begin{figure}[htp]
\sidecaption
\includegraphics[scale=1]{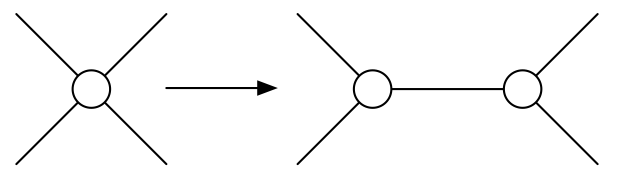}
\caption{Replacing a vertex $v$ by $(\deg(v)-2)$ vertices.}
\label{fig:vxreplace}
\end{figure}

We will now construct a tensor network whose value is the partition function $Z_G$.  Given the graph $G$, we define the vertices of  $G\pri=(V\pri,E\pri)$ as follows:
\be V\pri=V\cup\{v_e:e\in E\}\ee
We are simply adding a vertex for each edge $e$ of $G$, and identifying it by $v_e$.  We then define the edge set of $G\pri$:
\be E\pri=\{(x,v_e), (v_e,y):(x,y)=e\in E\}\ee
So, the end product $G\pri$ resembles the original graph; we have simply added vertices in the middle of each edge of the original graph.  The tensors that will be identified with each vertex will be defined separately for the vertex set $V$ of the original graph and the set $V_E=\{v_e\}$ of vertices we added to define $G\pri$.  In each case, they will be dimension $q$ tensors for a $q$-state model.  For $v\in V$, $M[v]$ is an `identity' operator;  that is, it takes on the value 1 when all indices are equal, and 0 otherwise:
\be (M[v])_{i_1,..,i_m}=\begin{cases}1&\text{if }i_1=i_2=...=i_m\\0&\text{otherwise}\end{cases}\ee
The vertices in $V_E$ encode the actual interactions between neighbouring particles.  The tensor associated with $v_e$ is
\be (M_{v_e})_{s,t}=e^{-\beta h_e(s,t)}\ee
Now, let us consider the value of the tensor network in terms of $q$-edge colourings of $G$:
\be\label{coloursum} T(G\pri,\mathbf{M})=\sum_\gamma\left(\prod_{v\in V\pri}M[v]^\gamma\right)\ee
where $\gamma$ runs over all $q$-edge colourings of $G\pri$.  Based on the definitions of the tensors $M[v]$ for $v\in V$, we see that $\prod_{v\in V\pri}M[v]^\gamma$ is non-zero only when the edges incident at each vertex are all coloured identically.  This restriction ensures that each non-zero term of the sum (\ref{coloursum}) corresponds to a configuration $\sigma$ of the $q$-state model.  That is, a configuration $\sigma$ corresponds to a $q$-edge colouring $\gamma$ of $G\pri$ where each $\gamma(e)=\sigma_v$ whenever $e$ is incident with $v\in V$.  This gives us the following equality:
\begin{flalign*}
T(G\pri, \mathbf{M})=&\sum_\gamma\left(\prod_{v\in V\pri}M[v]^\gamma\right)\\
=&\sum_\sigma\prod_{e=(u,v)} e^{-\beta h_e(\sigma_u,\sigma_v)}\\
=&\sum_\sigma e^{-\beta H(\sigma)}\\
=&Z_G(\beta)
\end{flalign*}
We now consider the time required to approximate the value of this tensor network.  Recall that the running time is given by
\be O\left(\text{poly}(q^{D(G\pri)})\cdot\size{V(G\pri)}\cdot\epsilon^{-1}\right)\ee
First, we observe that, if we assume that $G$ is connected, then $\size{V(G\pri)}$ is $O(\size{E})$.  We have not placed any restrictions on the maximum degree $D(G\pri)$.  The vertices of $G\pri$ of high degree must be in $V$, since all vertices in $V_E$ have degree 2.  Now, the tensors assigned to vertices of $V$ are just identity operators; they ensure that the terms of the sum from (\ref{coloursum}) are non-zero only when all the edges incident at $v$ are coloured identically.  As a result, we can replace each vertex $v\in V$ with $\deg(v)>3$ by $(\deg(v)-2)$ vertices, each of degree 3.  See figure \ref{fig:vxreplace} for an example.
The tensor assigned to each of these new vertices is the identity operator   of rank 3.  It is not difficult to show that this replacement does not affect the value of the tensor network.  It does, however, affect the number of vertices in the graph, adding a multiplicative factor to the running time, which can now be written
\be O\left(\text{poly}(q)\cdot D(G)\cdot\size{E}\cdot\epsilon^{-1}\right)\ee
For more detail on the approximation scale of this algorithm, the reader is referred to \cite{arXiv:0805.0040}.

Arad and Landau also discuss a useful restriction of $q$-state models called {\em difference models}.  In these models, the local Hamiltonians $h_e$ depend only on the difference between the states of the vertices incident with $e$.  That is, $h_e(\sigma_u,\sigma_v)$ is replaced by $h_e(\size{\sigma_u-\sigma_v})$, where $\size{\sigma_u-\sigma_v}$ is calculated modulo $q$.  Difference models include the well-known Potts, Ising and Clock models.  Arad and Landau show that the approximation scale can be improved when we restrict our attention to difference models.

Van den Nest et al. \cite{arXiv:0812.2368,arXiv:0812.2127} show that the partition function for difference models can be described as the overlap between two quantum states,
\be Z_G=\bra{\psi_G}\left(\bigotimes_{e\in E(G)}\ket{\alpha_e}\right).\ee
Van den Nest uses the state $\ket{\psi_G}$ to encode the structure of the graph, while each state $\ket{\alpha_e}$ encodes the strength of the local interaction at $e$.  The actual computation of this overlap is best described as a tensor network.  While this description does not yield a computational speedup over the direct method described above, it is an instructive way to deconstruct the problem.  In the case of planar graphs, it also relates the partition function of a graph $G$ and its planar dual.

Geraci and Lidar \cite{arXiv:0801.4833,arXiv:quant-ph/0703023} show that the Potts partition may be efficiently and exactly evaluated for a class of graphs related to irreducible cyclic cocycle codes.  While their algorithm does not employ tensor networks, it is interesting to consider the implications of their results in the context of tensor networks.  Is there a class of tensor networks whose value can be efficiently and exactly calculated using similar techniques?

\section{Conclusion}
In this chapter, we have reviewed quantum algorithms of several types:  those based on the quantum Fourier transform, amplitude amplification, quantum walks, and evaluating tensor networks.  This is by no means a complete survey of quantum algorithms; most notably absent are algorithms for simulating quantum systems.  This is a particularly natural application of quantum computers, and these algorithms often achieve an exponential speedup over their classical counterparts.

Algorithms based on the quantum Fourier transform, as reviewed in section \ref{sec:quantum-fourier-transform} include some of the earliest quantum algorithms that give a demonstrable speedup over classical algorithms, such as  the algorithms for Deutsch's problem and Simon's problem.  Many problems in this area can be described as hidden subgroup problems.  Despite the fact that problems of this type have been well-studied, many questions remain open.  For example, there is no efficient quantum algorithm known for most examples of non-Abelian groups.  In particular, the quantum complexity of the graph isomorphism problem remains unknown.

Similarly, the family of algorithms based on amplitude amplification and estimation have a relatively long history.  While Grover's original searching algorithm is the best-known example, this approach has been greatly generalized.  For example, in \cite{BHMT00} it is shown how the same principles applied by Grover to searching can be used to perform amplitude estimation and quantum counting.  Amplitude amplification and estimation have become ubiquitous tools on quantum information processing, and are often employed as critical subroutines in other quantum algorithms.

Quantum walks provide an interesting analogue to classical random walks.  While they are inspired by classical walks, quantum walks have many properties that set them apart.  Some applications of quantum walks are natural and somewhat unsurprising; using a walk to search for a marked element is an intuitive idea.  Others, such as evaluating AND-OR trees are much more surprising.  Quantum walks remain an active area of research, with many open questions.  For example, the relative computational capabilities of discrete and continuous time quantum walks is still not fully understood.

The approximation algorithm for tensor networks as described in this chapter \cite{arXiv:0805.0040} is relatively new.  However, it is inspired by the earlier work of Aharonov et al. \cite{arXiv:quant-ph/0511096}, as well as Yard and Wocjan \cite{arXiv:quant-ph/0603069}.  The tensor network framework allows us to capture a wide range of problems, particularly problems of a combinatorial nature.  Indeed, many $\#P$-hard problems, such as evaluating the Tutte polynomial, can be captured as tensor networks.  The difficulty arises from the additive nature of the approximation.  One of the most important questions in this area is when this approximation is useful, and when the size of the approximation window renders the approximation trivial.

WeÕve also briefly mentioned numerous other families of new quantum algorithms that we did not detail in this review article, and many others unfortunately go unmentioned. We have tried to balance an overview of some of the classic techniques with a more in depth treatment of some of the more recent and novel approaches to finding new quantum algorithms.

We expect and hope to see many more exciting developments in the upcoming years: novel applications of existing tools and techniques, new tools and techniques in the current paradigms, as well as new algorithmic paradigms.

\bibliographystyle{spmpsci}

\end{document}